\documentclass[namedreferences]{solarphysics}

\usepackage[hyperref,optionalrh]{spr-sola-addons} 
\usepackage{graphicx}        
\usepackage{color}           
\usepackage{breakurl}        
\usepackage{rotating}

\chardef\us=`\_

\begin{document}

\begin{article}
\begin{opening}

\title{Coronal Elemental Abundance: New Results from Soft X-ray Spectroscopy of the Sun}

\author[addressref={aff1},corref,email={kcshyama@ursc.gov.in}]{\fnm{Shyama}~\lnm{Narendranath}\orcid{0000-0001-9199-4925}}
\author[addressref={aff2},email={psreekumar@isro.gov.in}]{\fnm{P.}~\lnm{Sreekumar}}
\author[addressref={aff1}]{\fnm{Netra}~\lnm{S. Pillai}}
\author[addressref={aff3}]{\fnm{Singam}~\lnm{Panini}}
\author[addressref={aff1}]{\fnm{Sankarasubramanian}~\lnm{K.}}
\author[addressref={aff4}]{\fnm{Juhani}~\lnm{Huovelin}}

\address[id=aff1]{Space Astronomy Group, U R Rao Satellite Centre, ISRO, Bengaluru, India}
\address[id=aff2]{Indian Space Research Organisation HQ, Bengaluru, India}
\address[id=aff3]{Norwegian University of Science and Technology}
\address[id=aff4]{Division of Geophysics and Astronomy, Department of Physics, University of Helsinki, Helsinki, Finland}

\runningauthor{Narendranath et al.}
\runningtitle{Coronal elemental abundances}

\begin{abstract}

Elemental abundances in the solar corona are known to be different from those observed in the solar photosphere.  The ratio of coronal to photospheric abundance shows a dependence on the first ionisation potential (FIP) of the element. We  estimate FIP bias from direct measurements of elemental abundances from soft X-ray spectra using data from multiple space missions covering a range of solar activity levels. This comprehensive analysis shows clear evidence for a decrease in FIP bias around maximum intensity of the X-ray flare with coronal abundances briefly tending to photospheric values and a slow recovery as the flare decays. The departure from coronal abundances are larger for the low FIP elements Ca, Fe and Si than for S which have a mid FIP value. These changes in the degree of fractionation might provide inputs to model wave propagation through the chromosphere during flares.  

\end{abstract}
\end{opening}

\section{Introduction}
     \label{S-Introduction} 

The abundance of elements in the solar photosphere is well established from many decades of systematic observations of the Sun using a variety of techniques and this has resulted in high confidence photospheric abundance values. However, abundance studies in the solar corona and coronae of other stars, have not kept pace compared to photospheric studies. Though it is widely believed that there exists a steady flow of photospheric material into the corona, possibly channelled by structures associated with magnetic field configurations, the coronal abundance shows distinct deviations from photospheric values.
For the solar case, studies show enhanced abundance of elements with low first ionisation potential (FIP) over photospheric values, often categorized as a FIP bias (ratio of low FIP element coronal abundances to corresponding photospheric values). Stellar coronae also exhibit this effect \citep{1997ApJ...478..403D,1999ApJ...516..324L,2004HEAD....8.1003G} as well as the inverse termed as inverse-FIP (IFIP) effect (see Laming, 2015, for a review). Studies in the past such as those by \citeauthor{1992PhyS...46..202F} (1992), \citeauthor{1999A&A...348..286F} (1999), \citeauthor{2003SPD....34.1622P} (2003), \citeauthor{Feldman06} (2006), \citeauthor{2010ApJ...720.1721S} (2010), \citeauthor{Sylw12a} (2012), \citeauthor{2012ApJ...755...33S} (2012), \citeauthor{2014SoPh..289.1585N} (2014), \citeauthor{Dennis} (2015) and \citeauthor{Moore} (2018) to  name a few, suggest that this deviation or FIP bias is not the same for all elements. The range of reported values in the past, also suggests that FIP bias values may be nominally stable but possibly vary with solar cycle phase \citep{2017NatCo...8..183B} and spatial location on the Sun. 

%

\par The first result on the spectroscopic measurements of FIP bias variability during the evolution of a flare was provided by \citeauthor{1984Natur.310..665S} (1984). Ca abundances measured by the Rentgenovsky Spektrometr s Izognutymi Kristalami (RESIK) Bragg Crystal spectrometer were observed to increase as the flux in the narrow band continuum increased and showed a hysteresis during the decay of the flare. Flare to flare differences in Ca abundance were presented in further detail by \citeauthor{Sylwester_1998} (1998). 
\citeauthor{2014ApJ...786L...2W} (2014) measured the Fe abundances (from irradiance measurements) using the Extreme Variability Experiment (EVE) on board the Solar Dynamics Observatory (SDO) at the peak of strong flares (X and M) showing that the plasma has photospheric compositions. Measurements from the Flat Crystal Spectrometer (FCS) of the Soft  X-rays Polychromator on board the Solar Maximum Mission (SMM) showed large variations in abundances even within non-flaring active regions and bright points of the quiescent Sun \citep{saba94}.

\par Recent studies are providing evidence that suggest the key role of local magnetic fields in generating the observed FIP bias.  A detailed study of an active region by Baker et al. (2015) reports a decrease in FIP bias (abundances close to photospheric) that is linked to small scale evolution of the underlying magnetic field.  FIP bias in emerging flux regions in a coronal hole were studied by Baker et al. (2018) and it was shown that the changes were related to the magnetic topology of the regions. 
\citeauthor{2019ApJ...875...35B} (2019) observed IFIP patches at the foot points during two confined flares in an active region and 
 \citeauthor{Baker_2020} (2020) reported similar observations in a highly complex active region with multiple strong flares. Over solar cycle time scales, 
 \citeauthor{2017NatCo...8..183B} (2017) showed a strong correlation of Ne coronal to photospheric abundance ratios,  to the phases of the solar cycle. \citeauthor{2018JASTP.173...28P} (2018) further extended this to stars and have shown that the FIP bias variations are likely related to the large scale coronal magnetic field and activity level of the toroidal magnetic field. Such variations could also occur at shorter time scales and smaller spatial scales. \citeauthor {2017ApJ...844..153L} (2017) presented scenarios  where the dependence of the variations in FIP bias is linked to the origin and propagation sites of Alfvén waves.
 
\par Variations in coronal abundances in stellar flares have also been observed in quiescent coronae which are IFIP biased showing photospheric composition during those flares \citep{refId0,Laming_2009,2020AAS...23514807S}.

\par In this work, we provide clear evidence for FIP bias variation during the evolution of flares at short time scales (minutes to seconds). While our flare averaged abundance values are within the range reported earlier, this work systematically probes the variation of FIP bias of Si, S, Ca, and Fe, during the course of individual flares as well as their variations across flare classes. 

\section{Observations} \label{sec:obs}

\par Soft X ray spectroscopy of the solar corona is a useful diagnostic to remotely study the plasma composition and its changes over time. Crystal-based X-ray spectrometers have derived abundances using (line plus continuum) measurements in narrow spectral bands with high resolution and centered about the line of interest.  Ideally, X-ray spectra below 8 keV need to be measured simultaneously with the continuum to clearly derive plasma temperature and accurate estimates of X-ray line flux. Non-imaging broad band solar soft X-ray spectrometers have been flown as supporting instruments in many planetary missions since a knowledge of the rapidly evolving solar X-ray spectra is required to convert planetary X ray flux to elemental abundances. We used data from three missions with similar spectrometers (Si-PIN diodes) to derive elemental abundances of the corona during a range of solar activity levels.  

\subsection{SMART-1 XSM}

The X-ray Solar Monitor (XSM) on the Small Missions for Advanced Research in Technology-1 (SMART-1) (Huovelin et al., 2002) 
is a single pixel, non imaging Si-PIN detector with a 105$^{\circ}$ full field of view such that the full disk of the Sun is observed at nearly all times. The solar data from this mission spans from 03 March 2004 to 30 August 2006 which includes the Earth escape phase and lunar orbit phase when the solar activity was low. The useful operational energy range was $\approx$ 2 keV to 20 keV with some exceptions where the lower energy threshold comes down to 1.8 keV (Si abundances were derived only for such cases). We have used spectral data of nine long duration flares for this work which fall in the B-C class.

\subsection{Chandrayaan-1 XSM}

The X-ray Solar Monitor on Chandrayaan-1 is similar to the SMART-1 XSM except for a reduction in the aperture size to accommodate a wider dynamic range in the expected counts. This instrument measured the X-ray spectrum in the range of approximately 1.8 - 20 keV from several X-ray flares during the nine month mission life from 28 November 2008 to 29 August 2009 and the results were presented in Narendranath et al. (2014).
\citeauthor{2014SoPh..289.1585N} (2014).
Here we have used a C1 flare from XSM to look at the variation in FIP bias during the latter.

\subsection{MESSENGER- SAX}

The Solar X ray Assembly (SAX) onboard the Mercury Surface, Space Environment, Geochemistry and Ranging (MESSENGER) satellite (around Mercury) observed the Sun in X-rays during the period from 2004 to 2014. The observations at the distance of Mercury began in 2007. We have chosen 33 flares in the B-M class for this work which have a clean rise and decay profile. 
The spectral resolution of SAX is 600 eV at 5.9 keV with which the X-ray emission lines of Si, S and Ca are not well resolved. However as shown by Dennis et al. (2015),
 in the 1.6 to 4 keV region dominated by line emission, good fits do provide a measure of the elemental abundances.

\section{Analysis\label{sec:analysis}}
The XSMs measure solar spectra with a time resolution of 16 s while SAX spectra have typically 300 s binning. Since a range of activity levels are covered, we used different integration times, suitably adjusted for the strength of flares. The integrated flux  (photons/cm$^2$/s) in the 2 - 10 keV band multiplied by the integration time is taken as a measure of the strength of the flare for inter-flare comparison. 
 SAX has a flat spectral resolution of 600 eV across its energy range while energy dependent resolution is incorporated for the XSMs. On board radioactive sources help in the calibration of  XSMs and response matrices with temperature dependent corrections for each data set are generated. 

\par
\begin{figure}[h]    
                 
               \vspace{-0.3\textwidth}
               \hspace{0.015\textwidth}
               \includegraphics[width=0.53\textwidth,clip=]{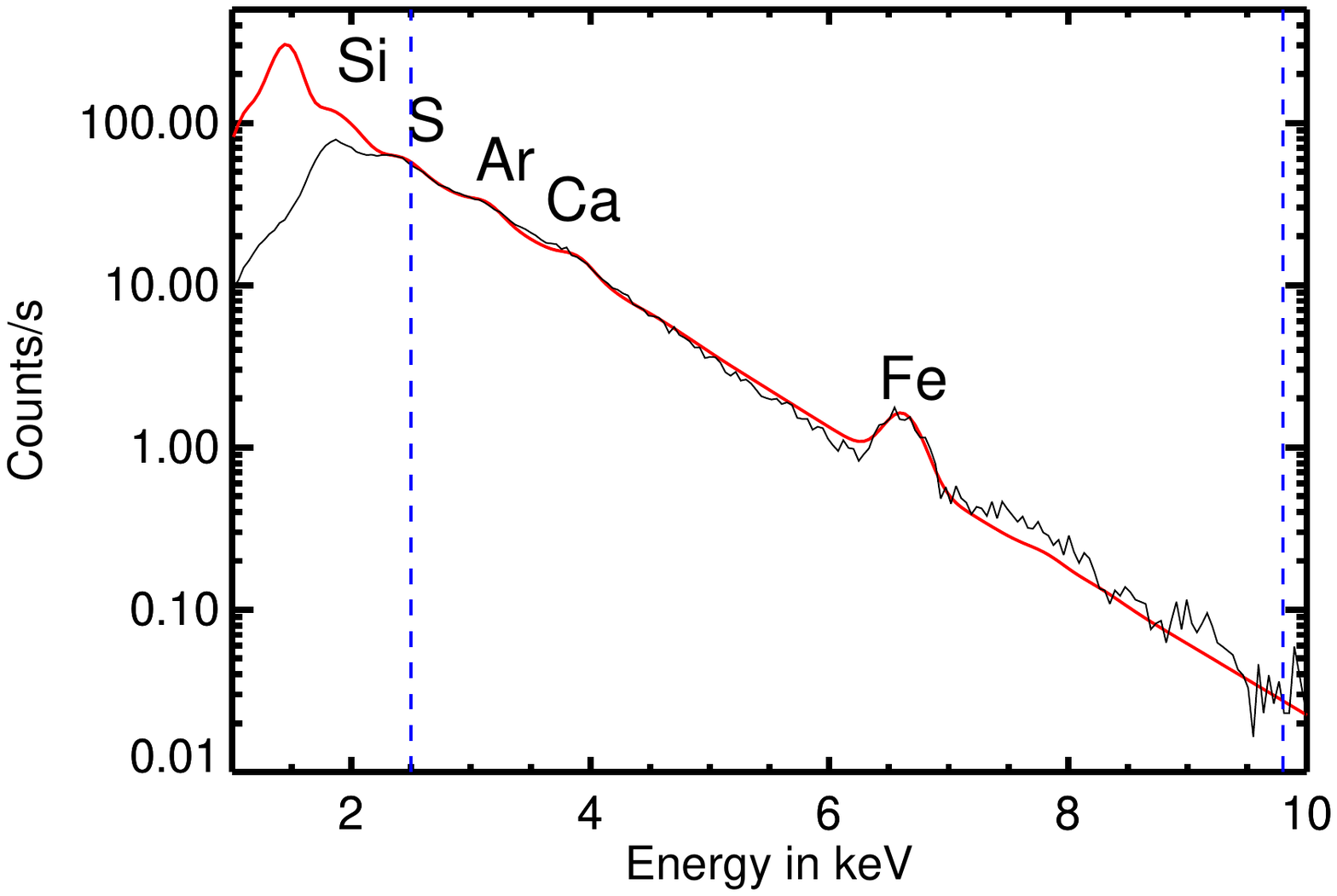}
               \hspace*{-0.03\textwidth}
               \includegraphics[width=0.53\textwidth,clip=]{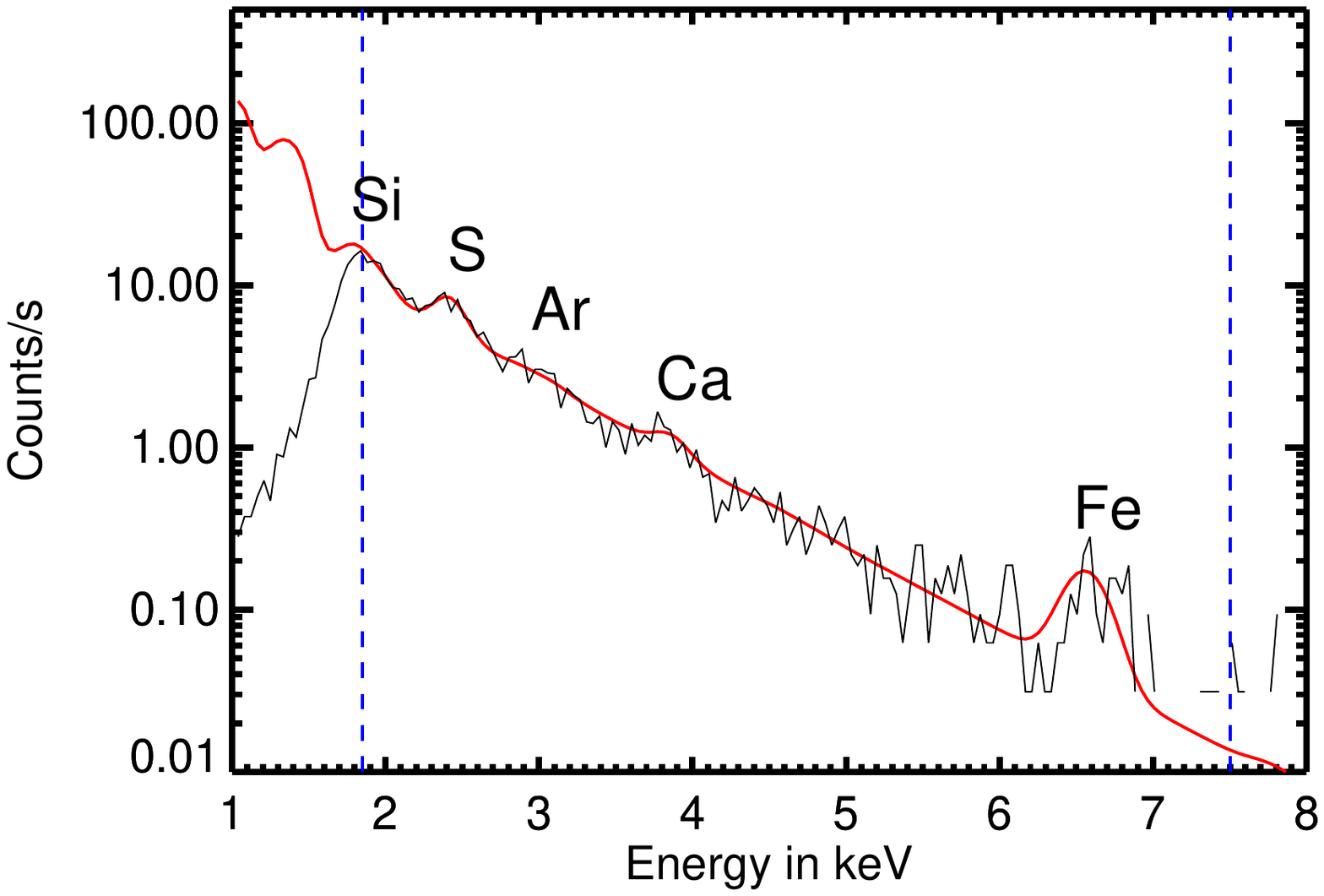}
                 
     \vspace{-0.25\textwidth}   
     \centerline{\Large \bf     
      \hspace{0.3 \textwidth}  \color{red}{(a)}
      \hspace{0.4\textwidth}  \color{red}{(b)}
         \hfill}
    \vspace{0.1\textwidth}    
   \centerline{\hspace*{0.015\textwidth}
               \includegraphics[width=0.6\textwidth,clip=]{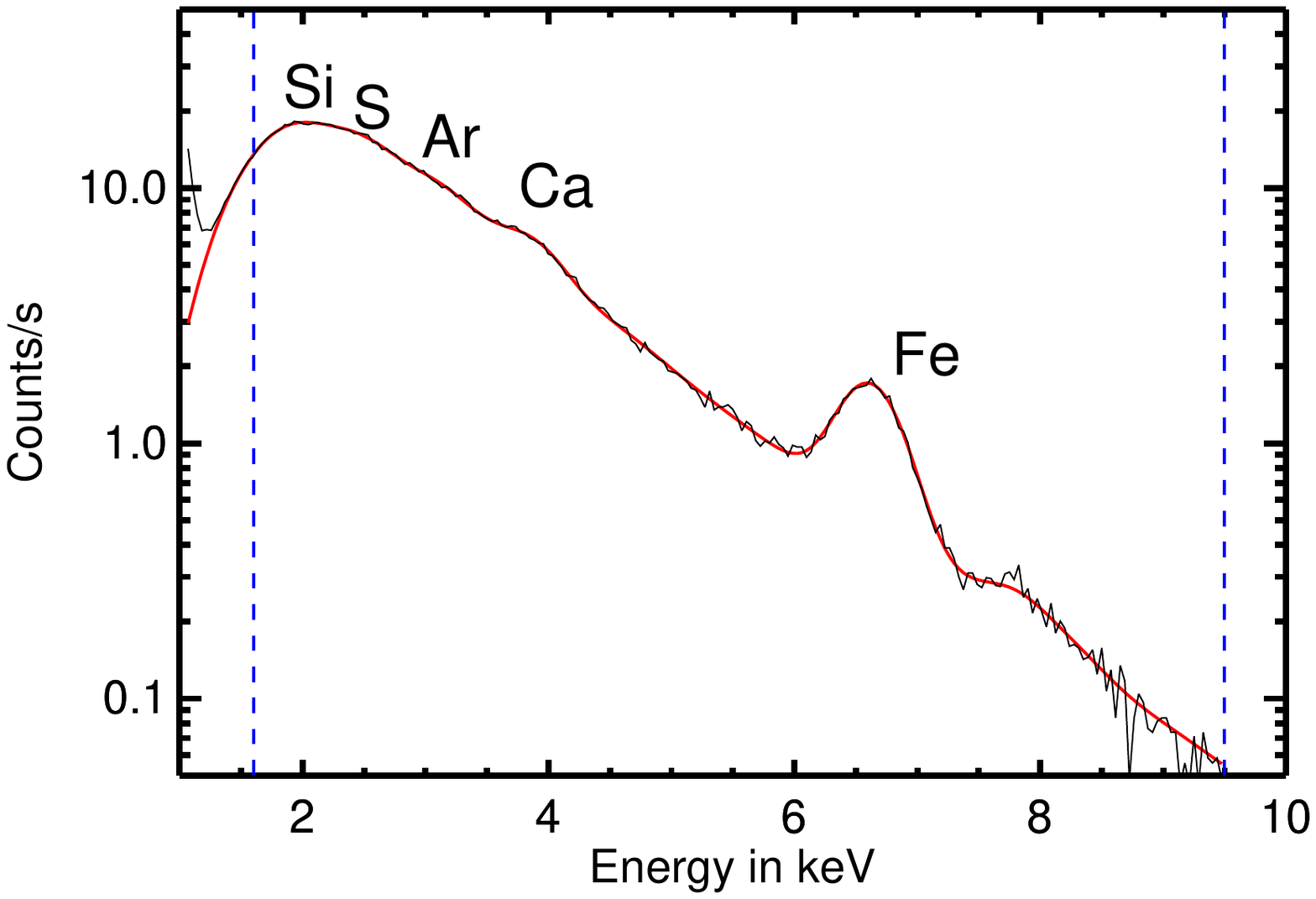}
               }
   
  \vspace{-0.6\textwidth}   
     \centerline{\Large \bf     
    \hspace{0.58 \textwidth} \color{red}{(c)}
         \hfill}
       \vspace{0.2\textwidth}  
              
\caption{  Representative spectral fits to flare spectrum from (a) SMART-1 XSM (26 June 2004) (b) Chandrayaan-1 XSM (5 July 2009) (c) MESSENGER SAX (2 January 2014). The red line indicates the best fit model (single temperature plasma varying the abundances) and the black line shows the data. Emission lines from elements are marked at expected line energies.  Vertical blue dotted lines indicate the energy range used for fitting the model to the spectrum.    }
   \label{spec}
   \end{figure}

 The spectral analysis was carried out with the Object Spectral Executive (OSPEX) package which comes as part of the standard solarsoft package (SSW) \cite{1998SoPh..182..497F}. The spectra were fitted with a thermal model (single temperature) calculated from CHIANTI (version 8.0.2) \cite{1997A&AS..125..149D} 
({\it vth$\_$abund} in OSPEX) which consists of a continuum and multiple emission lines. For larger flares, we added a non-thermal component to fit the continuum ({\it thick2} in OSPEX). Sample spectral fits from SMART-1 XSM, Chandrayaan-1 XSM and MESSENGER-SAX are shown in figure \ref{spec}. All three instruments are Si-PIN detectors that do not have any significant continuum background. In the absence of X-ray events, the background appears only as a peak at the lower channels. This is not the same for all detectors and varies with the leakage current. We have taken
intervals where the leakage current is less than 10 pA for all three detectors which ensures a more stable performance. The X-ray
emission measured is for the whole solar disc. We subtracted the pre-flare spectrum to
ensure that the derived abundances are indeed of the flare plasma. Elemental abundances were derived in the usual logarithmic ratio with respect to hydrogen (A$_{elem}$ = log(N$_{elem}$/N$_{H}$ + 12.0). The FIP bias values were derived with respect to the photospheric abundances as given in \citeauthor{2009ARAA..47..481A} (2009). 
Error bars are uncertainties arising from propagation of errors of fit parameters, while expressing it as a ratio to photospheric values (or FIP bias). The Fe line is excited only at higher plasma temperatures and is weak when temperatures are lower as in smaller flares.

\begin{table}
\centering
\caption{ Data source instruments \label{tab:inst}.}
\tiny
\label{T-simple}
\begin{tabular}{ccccc}     
  \hline                   
Instrument & Energy range & Spectral & Time & Obs.period \\
&&Resolution&Resolution& \\
SMART-1-XSM&1.8 to 20 keV&280 eV& 16 s&3/03/04 to 30/08/06 \\ \hline
Chandrayaan-1-XSM&1.8 to 20 keV&220 eV&16 s&28/11/08 to 29/08/09 \\ \hline
MESSENGER-SAX&1.6 to 9.5 keV&600 eV& 300 s& 2007 to 2014 \\ 

 \hline
\end{tabular}
\end{table}

\section{FIP Bias}\label{sec:fip}
FIP bias has been reported in the solar corona for several decades now. We have measured the abundances of low FIP elements Si, Ca and Fe, the mid-FIP element S and the high-FIP element Ar during 43 flares using full disk integrated spectra.  Individual flares which are  isolated events that dominate the soft X-ray emission over the entire solar disk, are analysed in detail.

\par
\begin{figure}[h]    
               \vspace{-0.4\textwidth}
               \hspace*{0.015\textwidth}
               \includegraphics[width=0.515\textwidth,clip=]{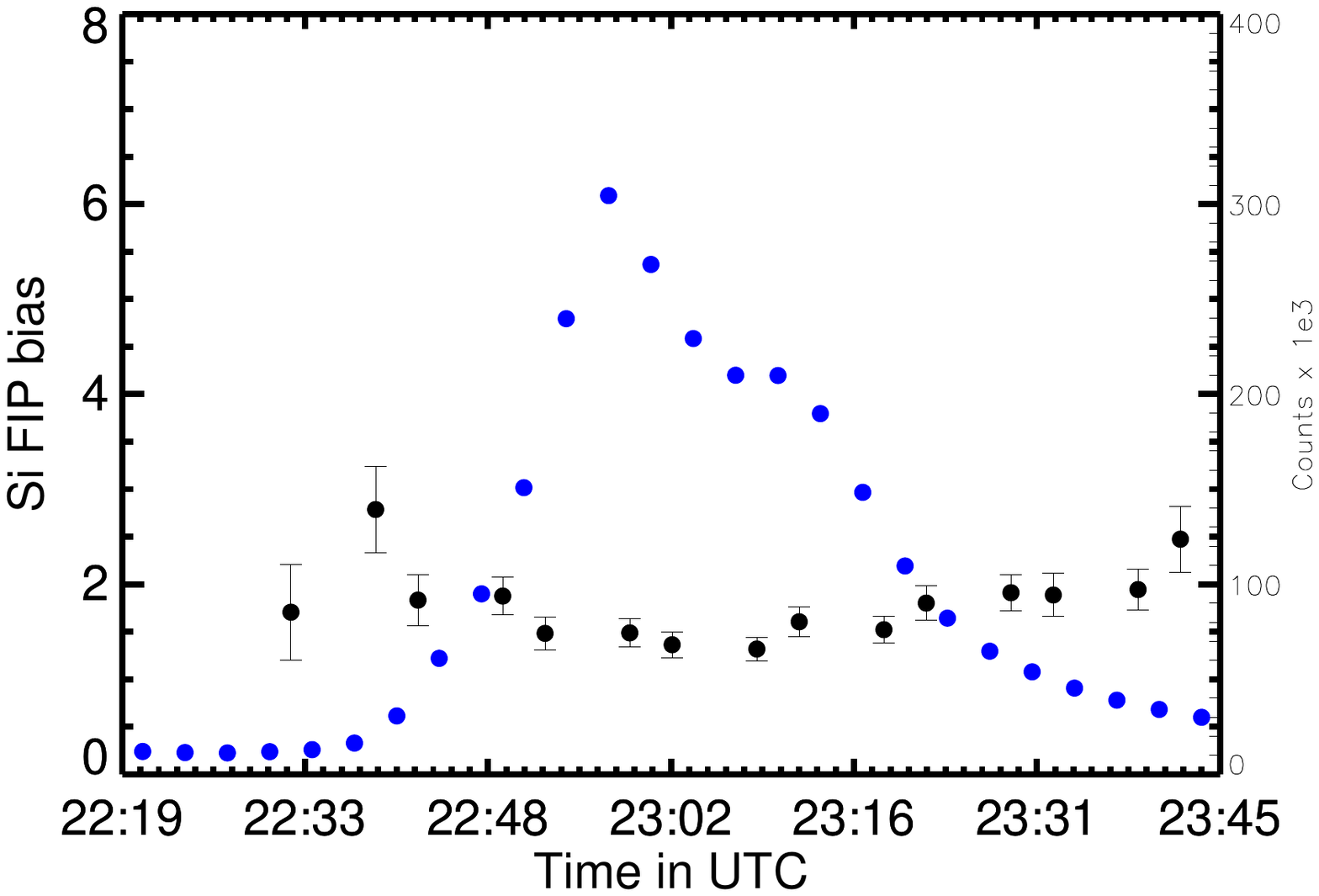}
              \hspace*{-0.03\textwidth}
               \includegraphics[width=0.515\textwidth,clip=]{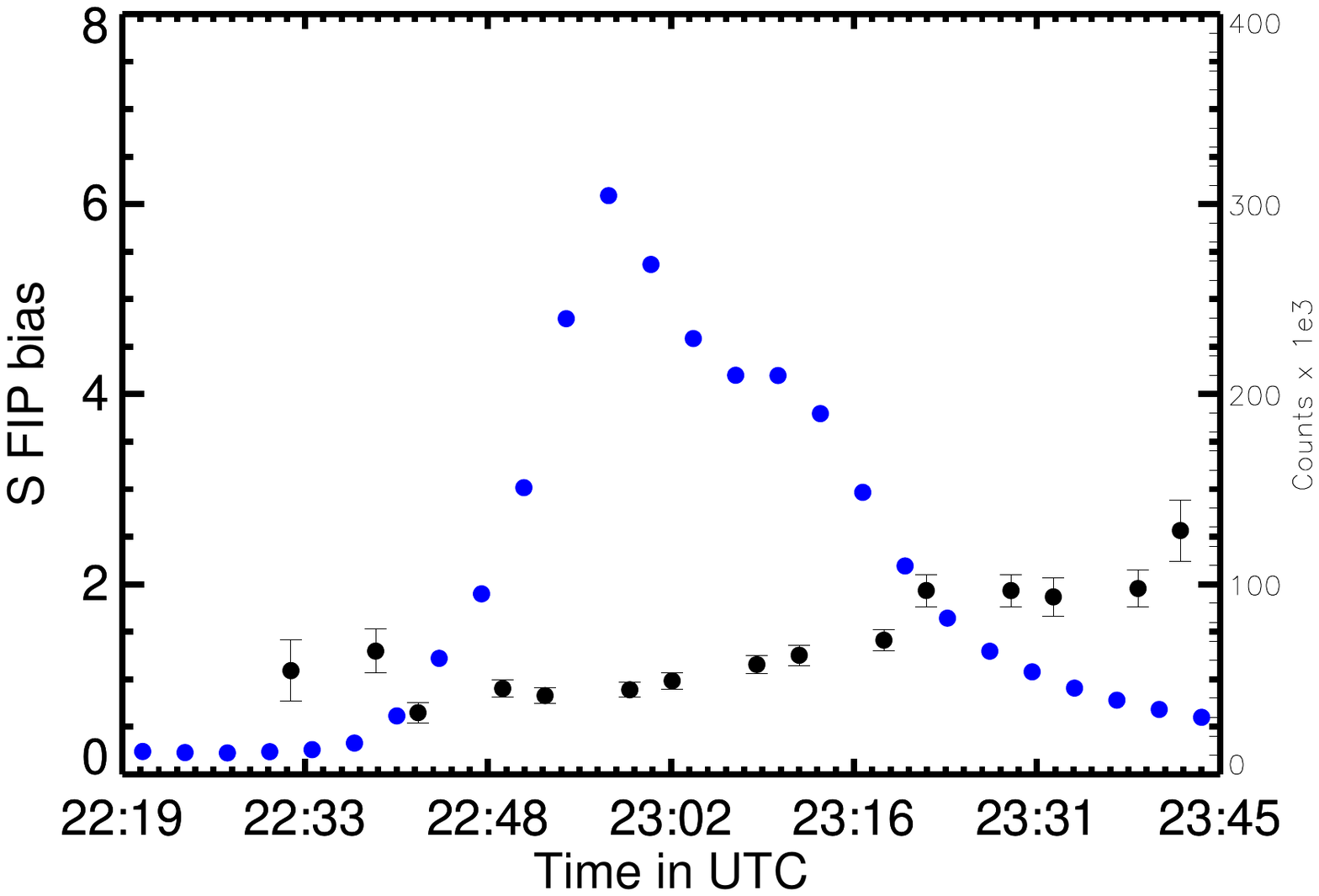}
              
     \vspace{-0.25\textwidth}   
     \centerline{\Large \bf     
      \hspace{0.1 \textwidth}  \color{red}{(a)}
      \hspace{0.4\textwidth}  \color{red}{(b)}
         \hfill}
    \vspace{-0.15\textwidth}    
   \centerline{\hspace*{0.015\textwidth}
               \includegraphics[width=0.515\textwidth,clip=]{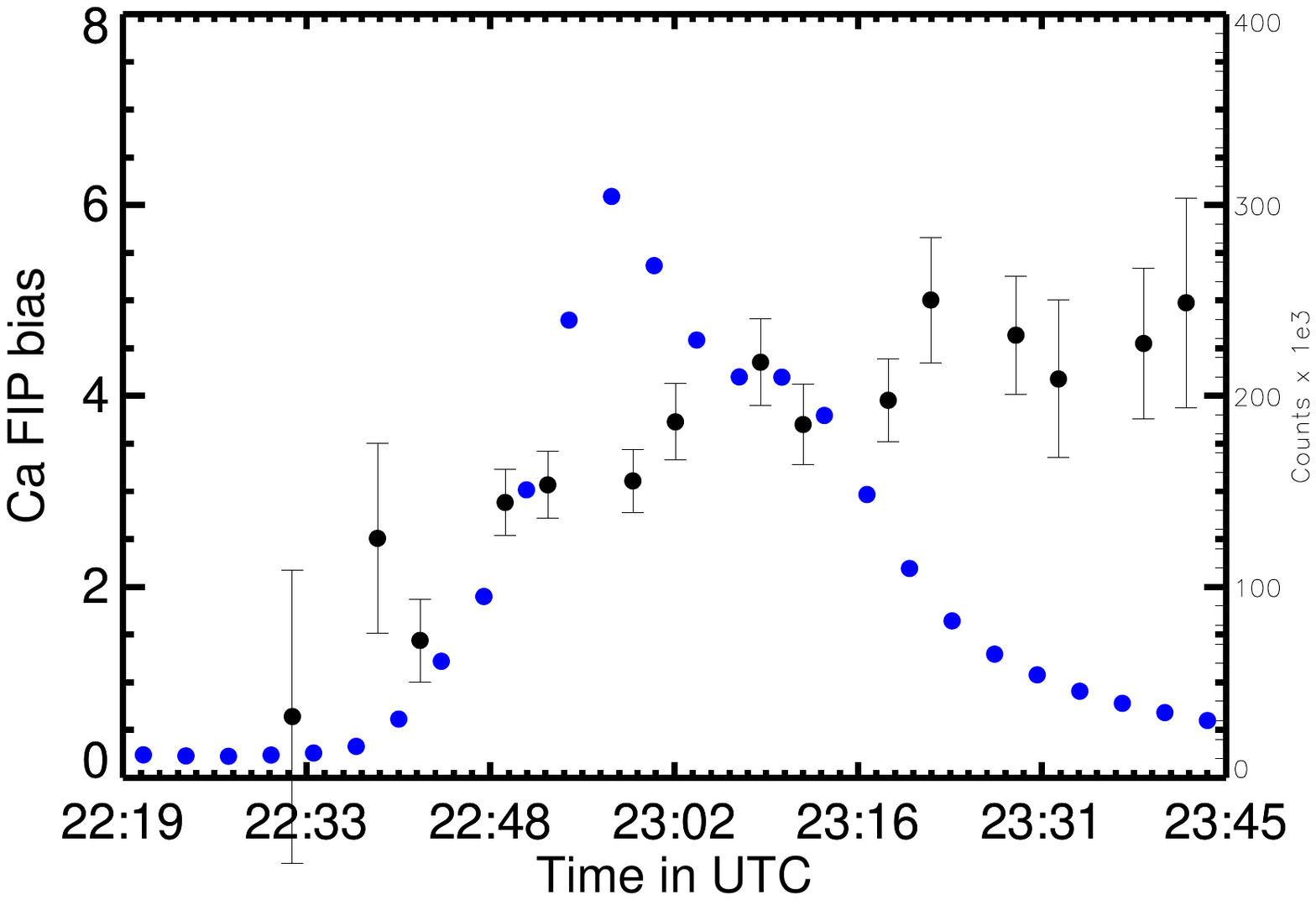}
               \hspace*{-0.03\textwidth}
               \includegraphics[width=0.515\textwidth,clip=]{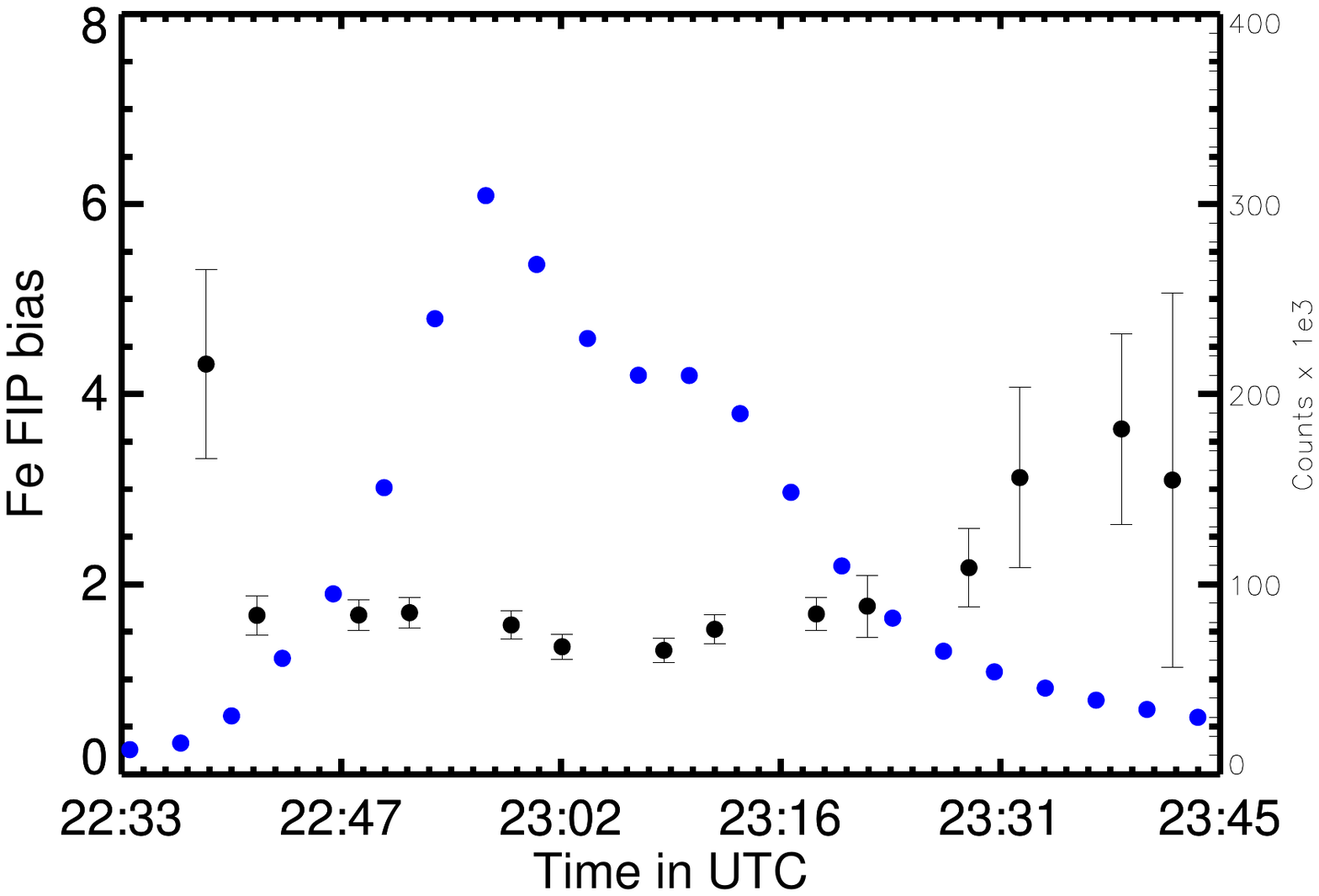}
              }
     \vspace{-0.25\textwidth}   
     \centerline{\Large \bf     
      \hspace{0.1 \textwidth} \color{red}{(c)}
      \hspace{0.4\textwidth}  \color{red}{(d)}
         \hfill}
    \vspace{0.2\textwidth}    
              
\caption{Variation of FIP bias values during an M flare on 6 January 2014. (a) Si, (b) S, (c) Ca, and (d) Fe. The light curve of the flare in the 2-10 keV energy range is plotted in blue. }
   \label{FIP_var_1}
   \end{figure}

\begin{figure}[h]    
               \vspace{-0.35\textwidth}
               \hspace*{0.015\textwidth}
               \includegraphics[width=0.52\textwidth,clip=]{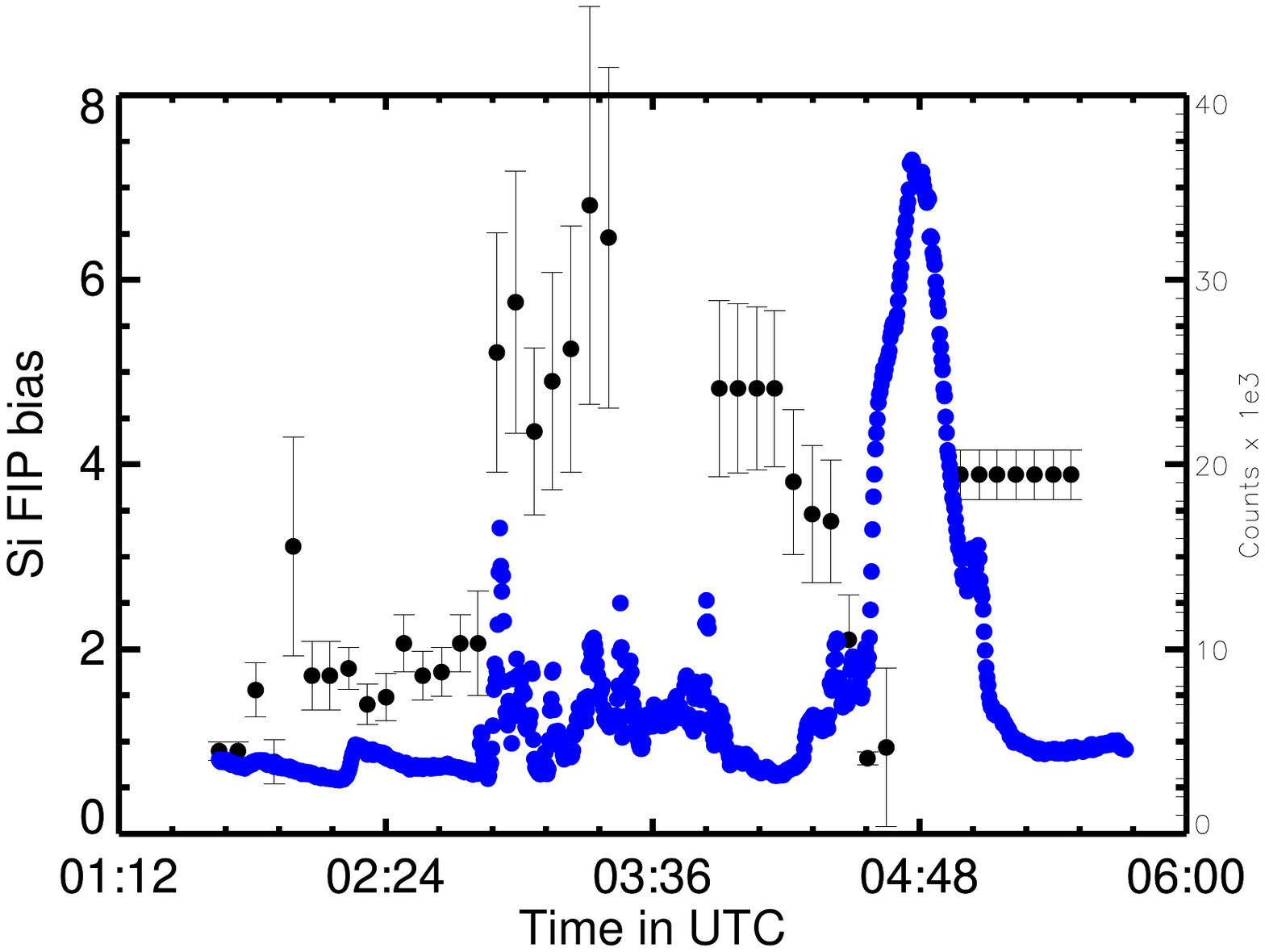}
               \hspace*{-0.03\textwidth}
               \includegraphics[width=0.52\textwidth,clip=]{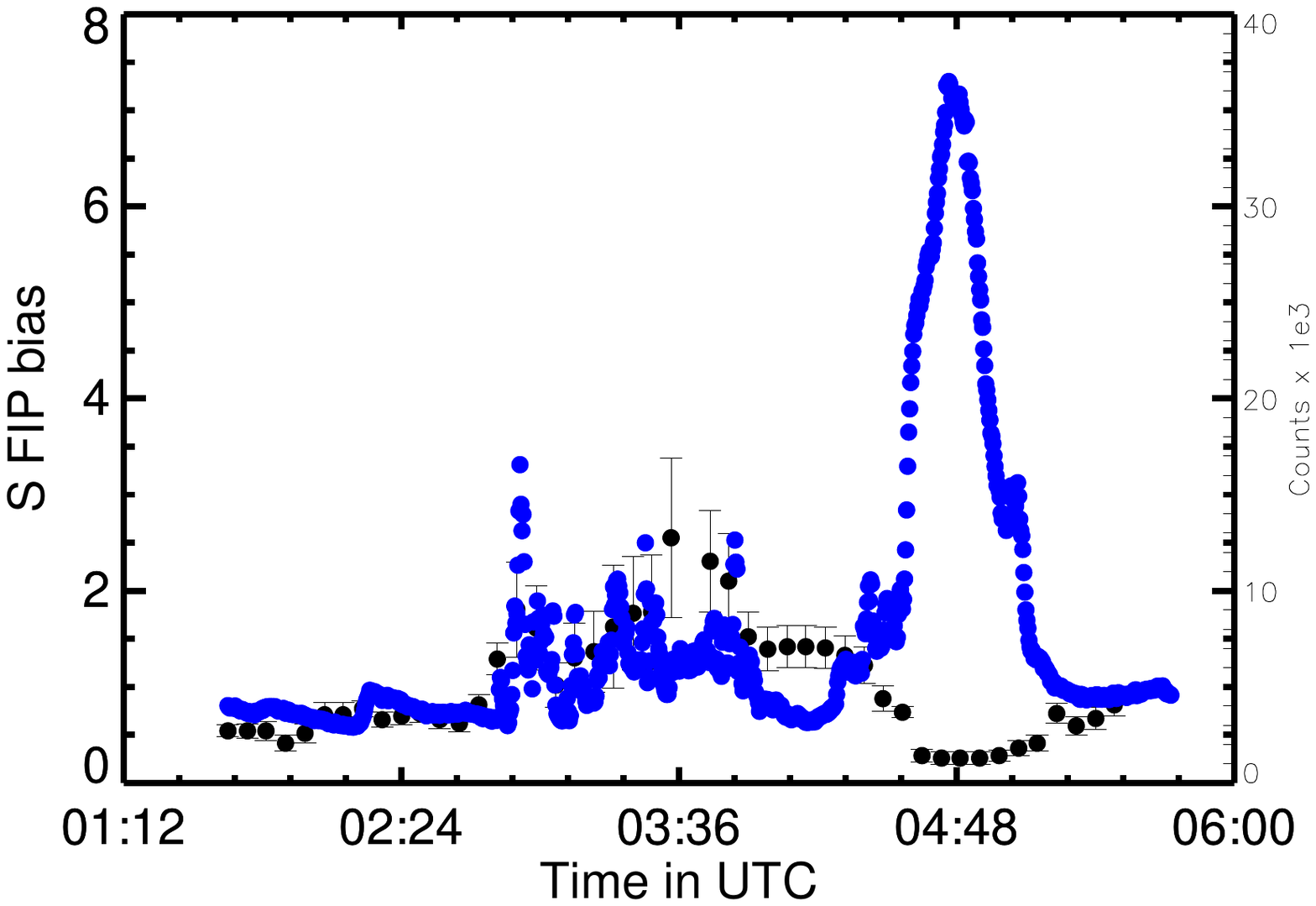}
              
     \vspace{-0.25\textwidth}   
     \centerline{\Large \bf     
      \hspace{0.1 \textwidth}  \color{red}{(a)}
      \hspace{0.415\textwidth}  \color{red}{(b)}
         \hfill}
   \vspace{-0.15\textwidth}    
                  
               \hspace*{0.015\textwidth}
               \includegraphics[width=0.52\textwidth,clip=]{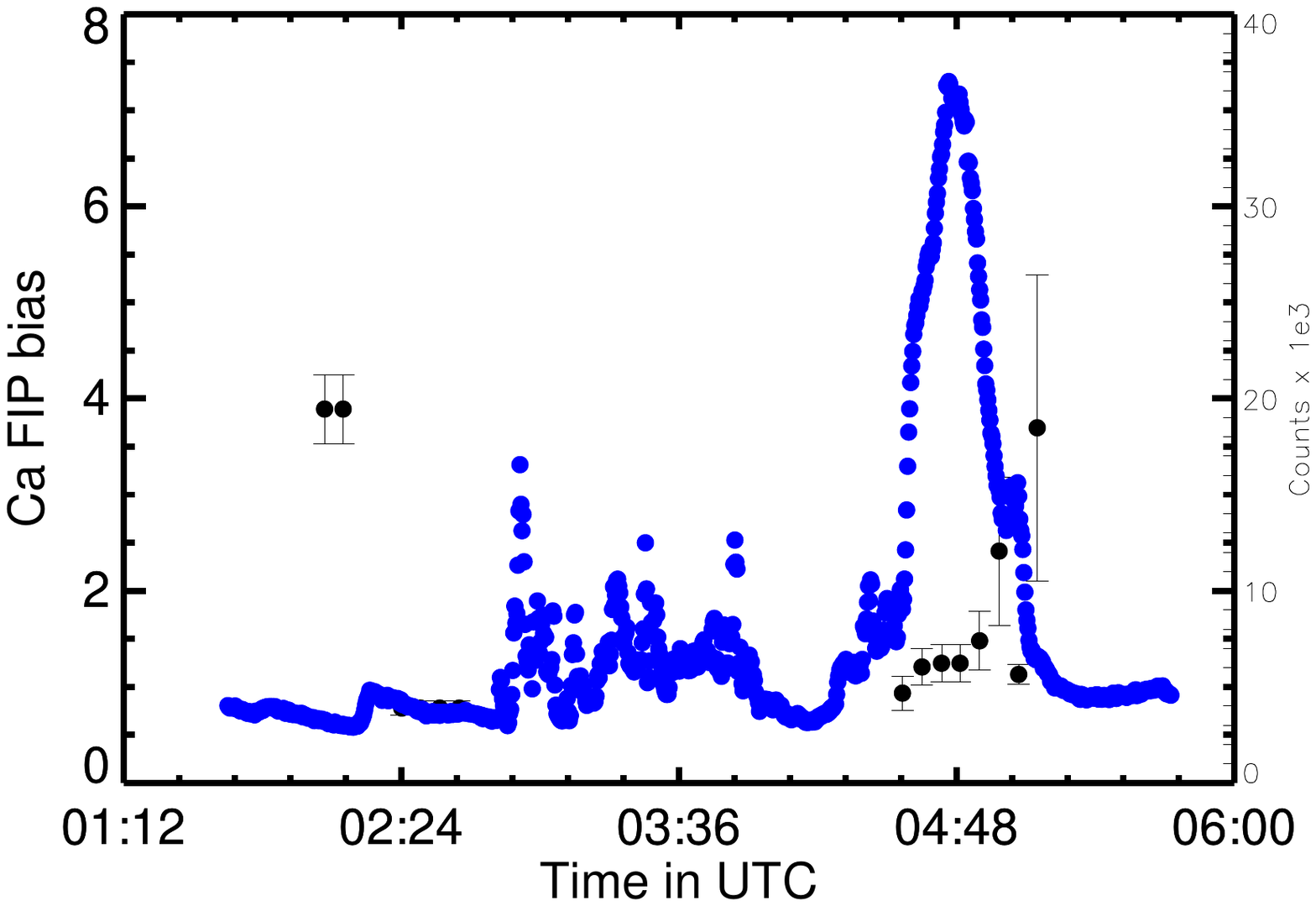}
               \hspace*{-0.03\textwidth}
               \includegraphics[width=0.52\textwidth,clip=]{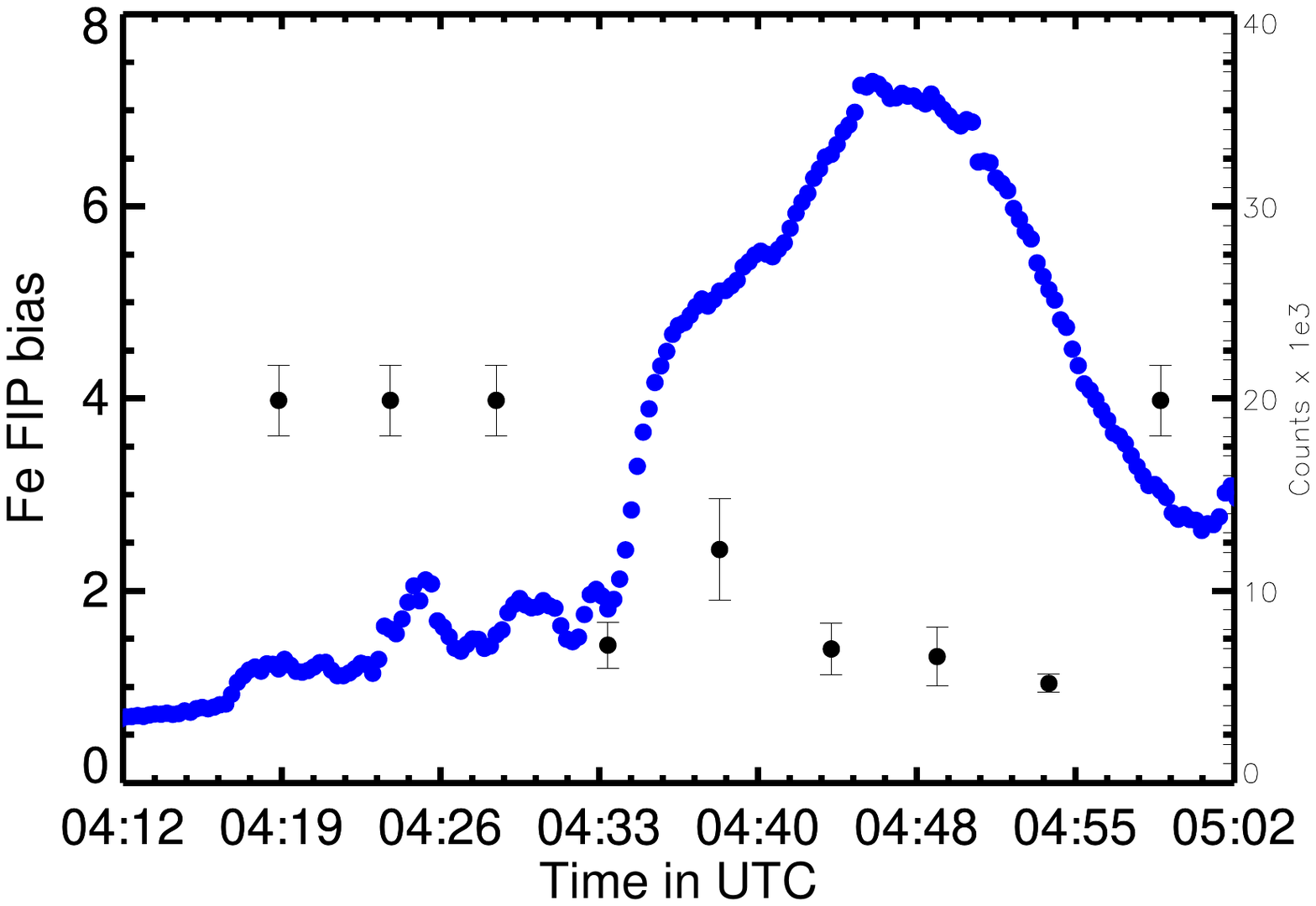}
              
     \vspace{-0.25\textwidth}   
     \centerline{\Large \bf     
      \hspace{0.1 \textwidth} \color{red}{(c)}
      \hspace{0.415\textwidth}  \color{red}{(d)}
         \hfill}
    \vspace{0.2\textwidth}    
              
\caption{Variation of FIP bias values during a B9 flare on 1 May 2004.(a) Si (b) S (c) Ca and (d) Fe. The light curve of the flare in the 2-10 keV energy range is plotted in blue. }
   \label{FIP_var_2}
   \end{figure}

\begin{figure}[h]    
   \centerline{\vspace*{-0.3\textwidth}
               \hspace*{0.015\textwidth}
               \includegraphics[width=0.52\textwidth,clip=]{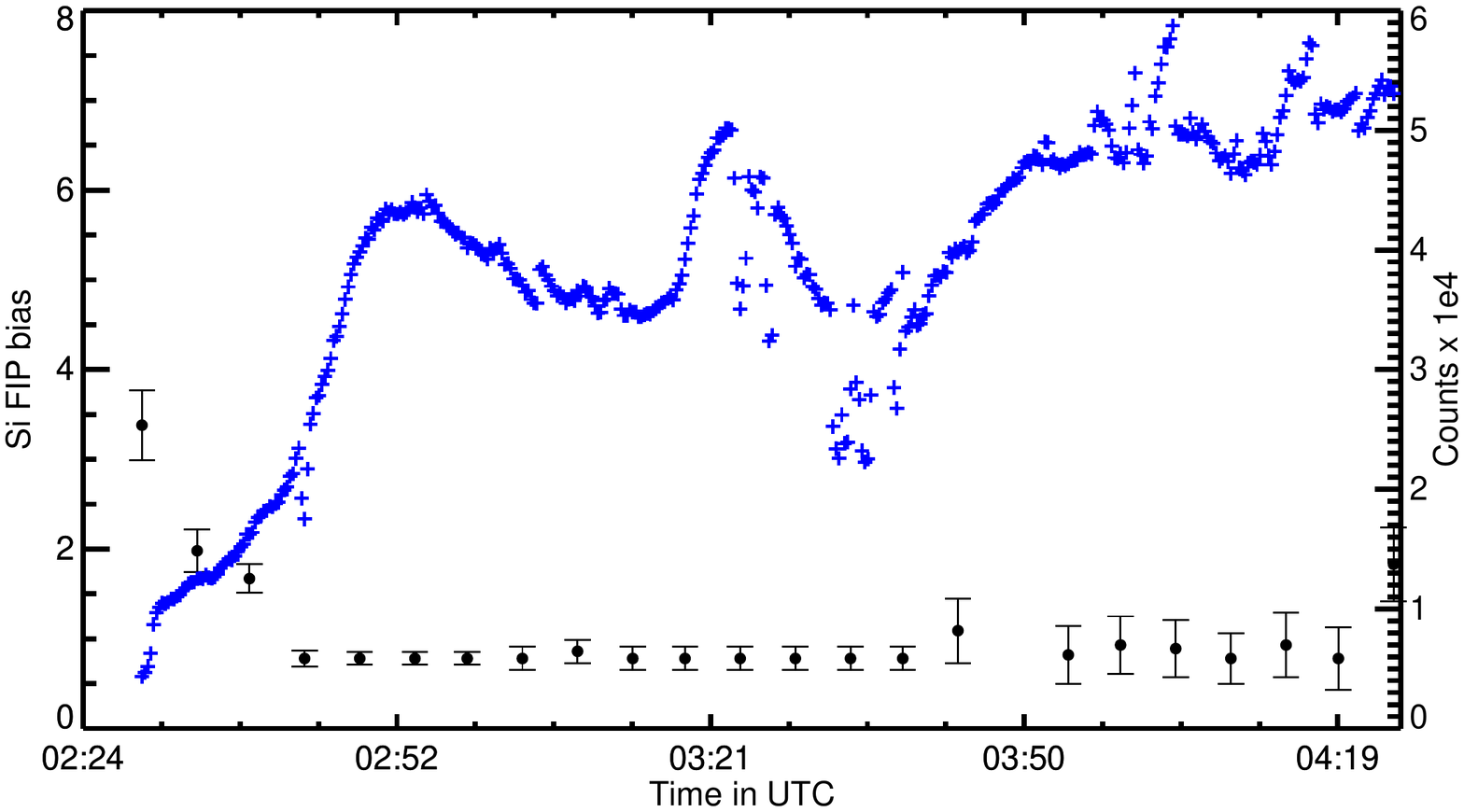}
               \hspace*{-0.03\textwidth}
               \includegraphics[width=0.52\textwidth,clip=]{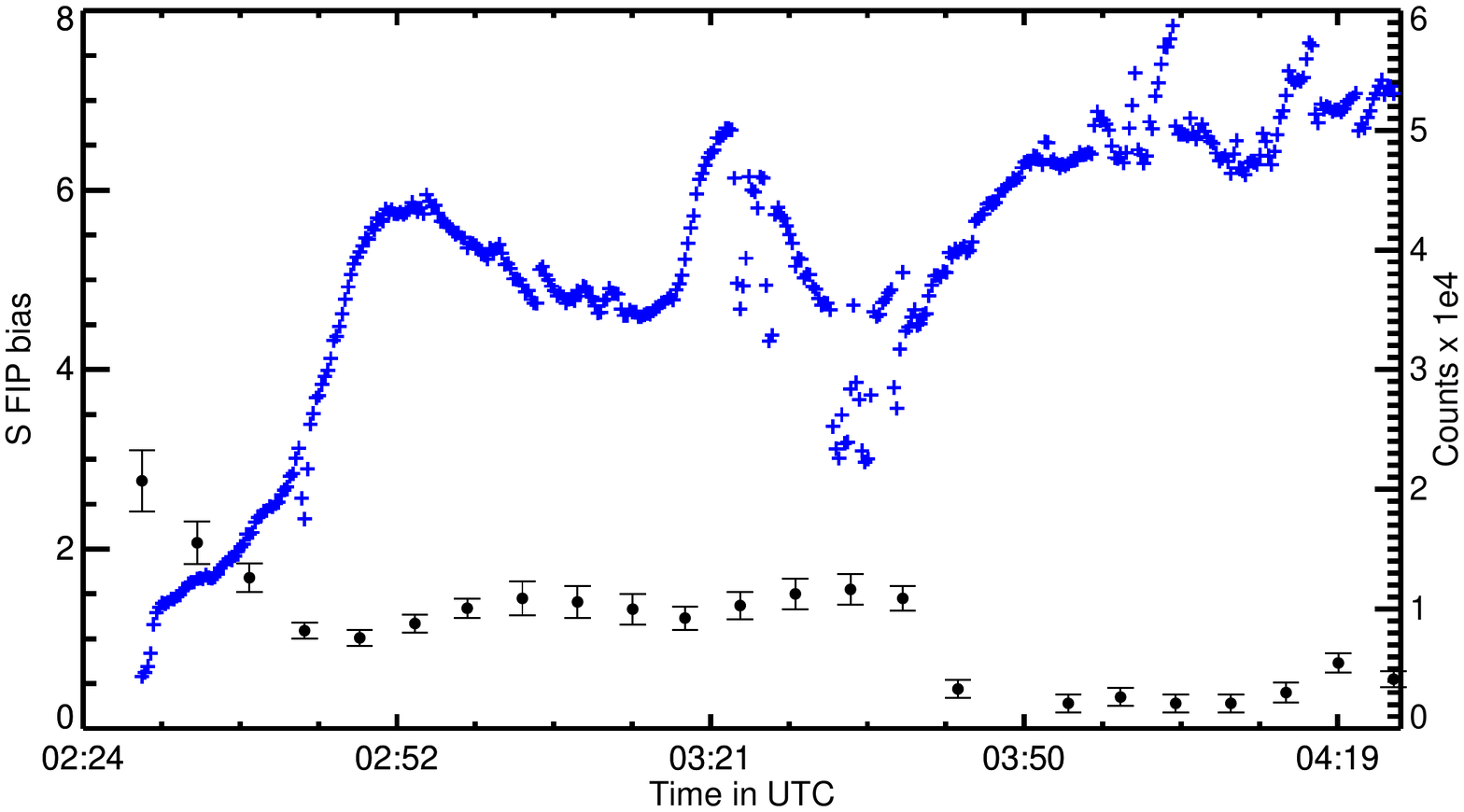}
              }
     \centerline{\Large \bf     
      \hspace{0.1 \textwidth}  \color{red}{(a)}
      \hspace{0.415\textwidth}  \color{red}{(b)}
         \hfill}
     \vspace{0.1\textwidth}    
   \centerline{\hspace*{0.015\textwidth}
               \includegraphics[width=0.52\textwidth,clip=]{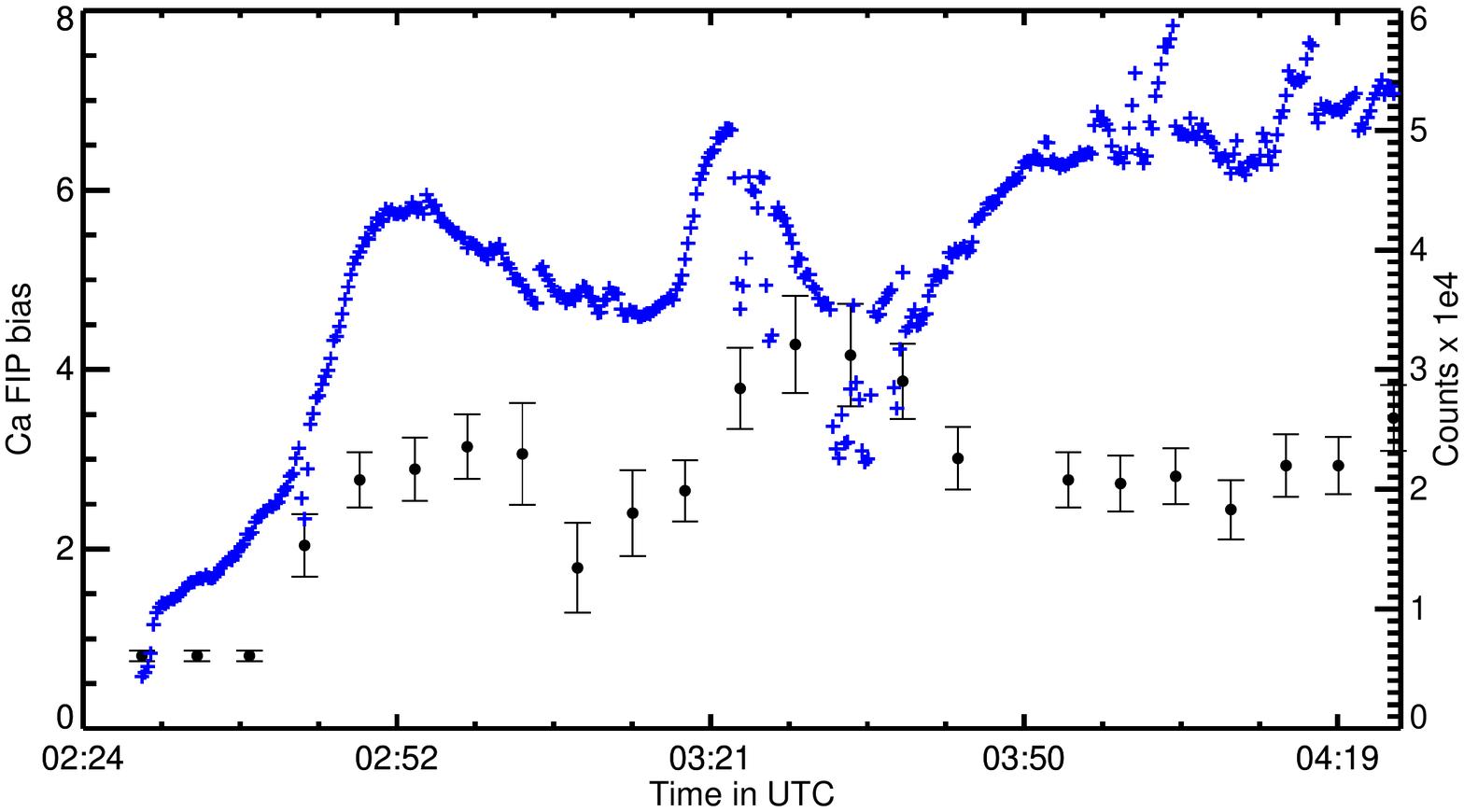}
               \hspace*{-0.03\textwidth}
               \includegraphics[width=0.52\textwidth,clip=]{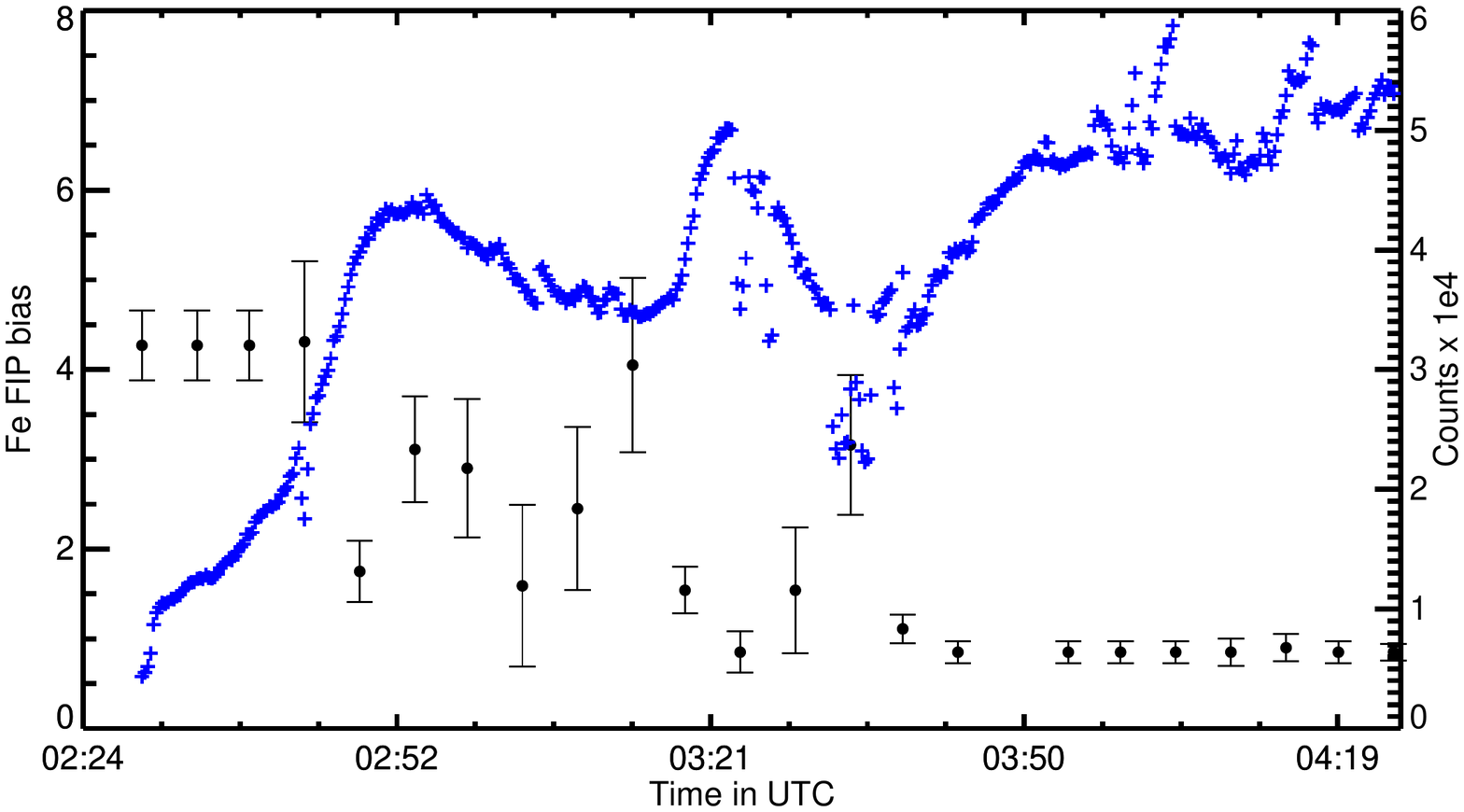}
              }
     \vspace{-0.3\textwidth}   
     \centerline{\Large \bf     
      \hspace{0.1 \textwidth} \color{red}{(c)}
      \hspace{0.415\textwidth}  \color{red}{(d)}
         \hfill}
     \vspace{0.2\textwidth}    
              
\caption{Variation of FIP bias values during a C flare on 2 January 2005.(a) Si, (b) S, (c) Ca, and (d) Fe. The light curve of the flare in the 2-10 keV energy range is plotted in blue. }
   \label{FIP_var_3}
   \end{figure}

 Figures \ref{FIP_var_1}, \ref{FIP_var_2}, and \ref{FIP_var_3} show the evolution of FIP bias values for each of the four low FIP elements as a function of time along with integrated light curves in the 2 - 10 keV range for three independent flares.  Ar abundances derived have higher uncertainties in comparison to the low FIP elements and hence are not shown here. The FIP bias value of low FIP elements systematically decreases as the flare intensity rises to a maximum and the bias value shows clear and consistent evidence for recovery as the flare decays. For example, in Figure ~\ref{FIP_var_1}d, Fe FIP bias decreases from around 4 to 1.4 and recovers to 4 after the flare. Si in Figure ~\ref{FIP_var_2}a decreases to 1 from a preflare value of about 5.
Ca FIP bias varies between 5 - 1 and does not show clear higher pre-flare values. 

The FIP bias of S varies from 2.5 during pre- and post-flares to 1 during the flare phase. These variations are significantly beyond the error bars within the observation interval but are less when compared to other elements. This behavior of FIP bias during a flare is independently shown by each of the low-FIP elements detected in the multiple spectra.
However, recovery to pre-flare values is not clearly visible in complex flaring events. For example, in Figure \ref{FIP_var_3} where this could arise due to sustained outburst lasting more than an hour with multiple re-connection events.  
 
\begin{sidewaystable}
\tiny
\resizebox{\textwidth}{!}{
\begin{tabular}{llccccccc}

Date&UTC&GOES&Temperature&Si&S&Ca&Fe&Flux (2-10 keV) \\
    &   &      class(approx)             &  (MK)              &  &  &  &  & in photons/cm$^2$/s \\
17.04.04&04:46:00 to 05:05:52&B3&9.46$\pm$0.48&0.40$\pm$0.26&0.29$\pm$0.12&1.12$\pm$0.85&4.27$\pm$0.39&192098.8    \\
01.05.04&04:30:32 to 05:40:40&B9&10.12$\pm$0.08& 3.89$\pm$0.27 &0.17$\pm$0.06 & 0.92$\pm$0.19 & 5.76$\pm$0.87 &  344738.7\\
05.05.04&19:27:04 to 20:04:24&B1&9.73$\pm$ 0.15 & 2.37$\pm$1.66 & 0.14$\pm$0.11  &  2.04 $\pm$   0.36 &   5.37 $\pm$   1.98 & 680821.6 \\
22.05.04&06:46:32 to 07:51:52&C1&9.47 $\pm$0.11& 3.75$\pm$0.43&    2.76 $\pm$0.33&    4.46$\pm$0.85&    8.53$\pm$1.50&  687424.6 \\
24.05.04&10:56:08 to 11:50:00&C3&14.76$\pm$0.04& 0.41$\pm$0.04 & 0.50$\pm$0.04 & 1.52$\pm$ 0.13 &   0.54$\pm$0.05& 2790486.5 \\
26.06.04&02:50:59 to 03:13:23&B2&7.80$\pm$0.05 &  0.39$\pm$0.06&    0.27$\pm$0.03&  0.92$\pm$0.16& 8.51$\pm$1.45&351884.7\\

05.07.04&05:14:11 to 05:41:39&A3&6.17$\pm$0.10&0.39$\pm$0.32&0.67$\pm$0.41&4.25$\pm$1.59&8.44$\pm$1.27&50387.25 \\
02.01.05&05:14:11 to 05:41:39&C1&10.37$\pm$0.32& 4.28$\pm$0.51&    1.74$\pm$0.28&    0.42$\pm$0.23&    8.53$\pm$1.71&  199968.3 \\

02.05.05&05:23:28 to 05:54:46&C&10.38$\pm$0.05&0.39$\pm0.05$&0.14$\pm$0.02&1.59$\pm$0.13&0.55$\pm$0.07&2833431.5  \\

05.07.09&08:58:05 to 09:23:09&A8&4.96$\pm$0.10&    3.89$\pm$0.27&    0.81$\pm$0.10&    1.54$\pm$0.61&    4.27$\pm$0.39&  40390.7  \\
02.01.14&01:52:44 to 06:57:24&M1&12.70$\pm$0.14&2.01$\pm$0.19&    1.41$\pm$0.12&    4.31$\pm$0.39& 3.12$\pm$    0.36& 2187142.0\\
04.01.14&10:04:47 to 12:28:07&M&13.09$\pm$0.09&  2.87$\pm$0.28&    1.70$\pm$    0.16&    3.78$\pm$0.41&    3.06$\pm$    0.31 &717274.6 \\
04.01.14&21:48:07 to 00:01:27&M1&15.38 $\pm$   0.05&    1.93$\pm$    0.16&    1.37$\pm$    0.11&    4.56$\pm$    0.36& 1.83$\pm$0.17&  5996155.5  \\
17.01.14&03:32:57 to 06:29:37&C2&10.23$\pm$    0.16&   3.77$\pm$    0.41&    2.40$\pm$    0.25&    3.97$\pm$    0.54&   8.38$\pm$ 1.38& 568340.4\\
17.01.14&18:59:37 to 21:19:37&C8&12.18 $\pm$  0.33&   1.96$\pm$    0.33&    1.80$\pm$    0.21 &   4.02$\pm$    0.44&   3.52$\pm$  0.67& 2290718.3 \\
18.01.14&02:16:29 to 03:13:09&B7&12.03 $\pm$   0.16&    2.69$\pm$    0.26&    2.08$\pm$    0.20&     3.44$\pm$    0.48& 5.42$\pm$    0.83&  1316521.5 \\
18.01.14&03:23:09 to 04:19:49&C1&10.13 $\pm$   0.12&   2.64$\pm$    0.26&    2.00$\pm$    0.19&     1.89$\pm$    0.45& 8.39$\pm$    1.36&  618368.5 \\
18.01.14&04:39:49 to 06:19:49&C&12.92$\pm$    0.12&   2.39$\pm$    0.21&    1.34$\pm$    0.12&     2.55$\pm$    0.33&  4.77$\pm$    0.57& 1925846.6 \\
18.01.14&10:40:29 to 11:53:49&C2&9.97$\pm$    0.10&   2.70$\pm$    0.32&    1.83$\pm$    0.21&     3.01$\pm$    0.51&  8.48$\pm$    1.31&   597014.6 \\
18.01.14&13:13:49 to 16:40:29&-&12.10$\pm$    0.31&   4.23$\pm$    0.40&    2.39$\pm$    0.24&     3.14$\pm$    0.48&  8.48$\pm$    2.07&  642723.6  \\
18.01.14&19:40:29 to 21:03:49&C1&9.56 $\pm$   0.13&   4.35$\pm$  0.44&    2.71$\pm$    0.27&     3.34$\pm$    0.60&  8.52$\pm$    1.53&   417767.7\\
20.01.14&02:23:55 to 03:47:15&C2&9.31$\pm$    0.13&   2.53$\pm$  0.23&    2.12$\pm$    0.19&     3.66$\pm$    0.56&  8.18$\pm$    1.60&  413894.8 \\
20.01.14&06:30:35 to 07:57:15&C1&9.58$\pm$    0.13&    3.22$\pm$  0.31&    2.12$\pm$    0.20&     1.55$\pm$    0.41&     8.53$\pm$    1.51&   402487.6 \\
20.01.14&14:41:15 to 16:37:65&-&11.66 $\pm$   0.10&   3.49$\pm$  0.29&    2.48$\pm$    0.23&     5.13$\pm$    0.51&     5.42$\pm$    0.65& 1603430.0 \\
20.01.14&21:41:15 to 00:04:35&C3&9.77$\pm$    0.13&    4.62$\pm$    0.59&  2.76$\pm$    0.38&     4.73$\pm$    0.87&     8.47$\pm$    1.21& 1089878.4 \\
27.01.14&02:14:21 to 03:31:01&-&9.47 $\pm$   0.11&   3.75$\pm$    0.43&    2.76$\pm$   0.33&     4.46$\pm$    0.85&     8.53$\pm$    1.50&   687424.6  \\
27.01.14&07:54:21 to 09:21:01&C4&11.52$\pm$    0.10&    3.19$\pm$ 0.28&    1.66$\pm$    0.15&     4.80$\pm$    0.54&     7.48$\pm$    0.94& 1317998.3\\
27.01.14&12:54:21 to 14:17:41&-&9.11$\pm$    0.15&   3.09 $\pm$  0.36&    2.17$\pm$    0.23&     3.15$\pm$    0.70&     8.44$\pm$    2.11&   281918.9 \\
27.01.14&18:28:21 to 20:18:21&C8&13.32$\pm$    0.24&    4.00$\pm$ 0.47&    2.75$\pm$    0.25&     6.97$\pm$    0.71&     3.53$\pm$    0.50& 1445175.0 \\
27.01.14&20:25:01 to 21:35:01&C5&12.28$\pm$    0.18&   2.92$\pm$ 0.33&    2.22$\pm$    0.20&     6.17$\pm$    0.63&     3.48$\pm$    0.48&  1644204.9 \\
27.01.14&22:12:58 to 23:57:58&M4&13.46$\pm$    0.76&    1.86$\pm$ 0.56&    1.63$\pm$    0.33&     4.55$\pm$    0.55&     2.37$\pm$    0.68& 3465951.0 \\
28.01.14&07:12:00 to 08:24:33&M3&20.12$\pm$    0.83&    6.40$\pm$    0.81&    2.73$\pm$    0.32&     5.84$\pm$    0.83&     3.10$\pm$    0.54& 2514868.8 \\
28.01.14&21:48:55 to 23:88:55&M1&15.04$\pm$    0.24&    2.56$\pm$    0.27&    1.43$\pm$    0.15&     4.21$\pm$    0.39&     2.14$\pm$    0.24&   4676428.0 \\
02.02.14&01:04:38 to 01:34:38&C6&13.00$\pm$    0.24&    3.49$\pm$    0.46&    2.28$\pm$    0.23&     4.99$\pm$    0.62&     4.62$\pm$    0.67& 1625441.1 \\
02.02.14&02:50:38 to 03:33:58&C2&11.02$\pm$    0.17&    2.04$\pm$    0.29&    1.87$\pm$    0.21&     4.05$\pm$    0.58&     4.98$\pm$    0.87&  938765.5 \\
02.02.14&08:00:38 to 09:07:18&M1&16.39 $\pm$   0.46&    3.50$\pm$    0.59&    2.60$\pm$    0.30&     5.81$\pm$    0.55&     1.42$\pm$    0.17& 7493439.5  \\
02.02.14&11:29:17 to 12:35:57&C9&12.75$\pm$    0.43&    3.21$\pm$    0.76&    2.70$\pm$    0.41&     3.94$\pm$    0.83&     3.53$\pm$    0.69& 1263622.6  \\
02.02.14&14:02:37 to 15:25:57&C8&13.39 $\pm$   0.31&    5.15$\pm$    0.70&    2.76$\pm$    0.30&     5.46$\pm$    0.69&     4.36$\pm$    0.65& 1790274.3     \\
02.02.14&16:09:17 to 17:29:17&M&12.92 $\pm$   0.34&    4.88$\pm$    0.77&    2.76$\pm$    0.34&     6.20$\pm$    0.78&     5.07$\pm$    0.81& 1409684.5 \\

06.02.14&03:56:47 to 05:20:07&C6&12.10$\pm$    0.10&   1.74$\pm$    0.16&    1.25$\pm$    0.10&     2.98$\pm$    0.31&     3.10$\pm$    0.34&  1856722.6 \\
07.02.14&04:43:27 to 05:26:47&M1&14.54 $\pm$   0.13&    1.94$\pm$    0.19&    1.76$\pm$    0.15&     3.99$\pm$    0.36&     1.67$\pm$    0.17&  5151088.5  \\
09.02.14&15:28:04 to 17:34:44&M&12.31 $\pm$   0.10&    2.01$\pm$    0.18&    1.60$\pm$    0.13&     3.45$\pm$    0.31&    2.42$\pm$    0.25& 2960934.5 \\                                                             
12.02.14 &00:05:59 to 00:06:46&-&8.86 $\pm$   0.13&    2.44$\pm$    0.24&    1.72$\pm$    0.18&     3.20$\pm$    0.66&     8.30$\pm$    1.77& 393868.4 \\

09.09.14&00:10:30 to 03:10:30&M4&11.12$\pm$    0.09&    1.83$\pm$    0.16&    1.57$\pm$    0.13&     3.92$\pm$    0.38 &    4.67$\pm$  0.55& 1228230.1 \\

\end{tabular}}
\caption{\label{tab:Bias}List of solar flares analysed in this study. The FIP bias values are given for all the low FIP elements in the spectral range studied. The flux is the integrated flux in the 2 to 10 keV band. Some of the flares seen in SAX MESSENGER were not observed in GOES and hence the GOES class not mentioned for those cases.}
\end{sidewaystable}

\par
\begin{figure}[h]    
        \vspace{-0.05\textwidth}
         \centerline{\hspace*{0.015\textwidth}
               \includegraphics[width=0.52\textwidth,clip=]{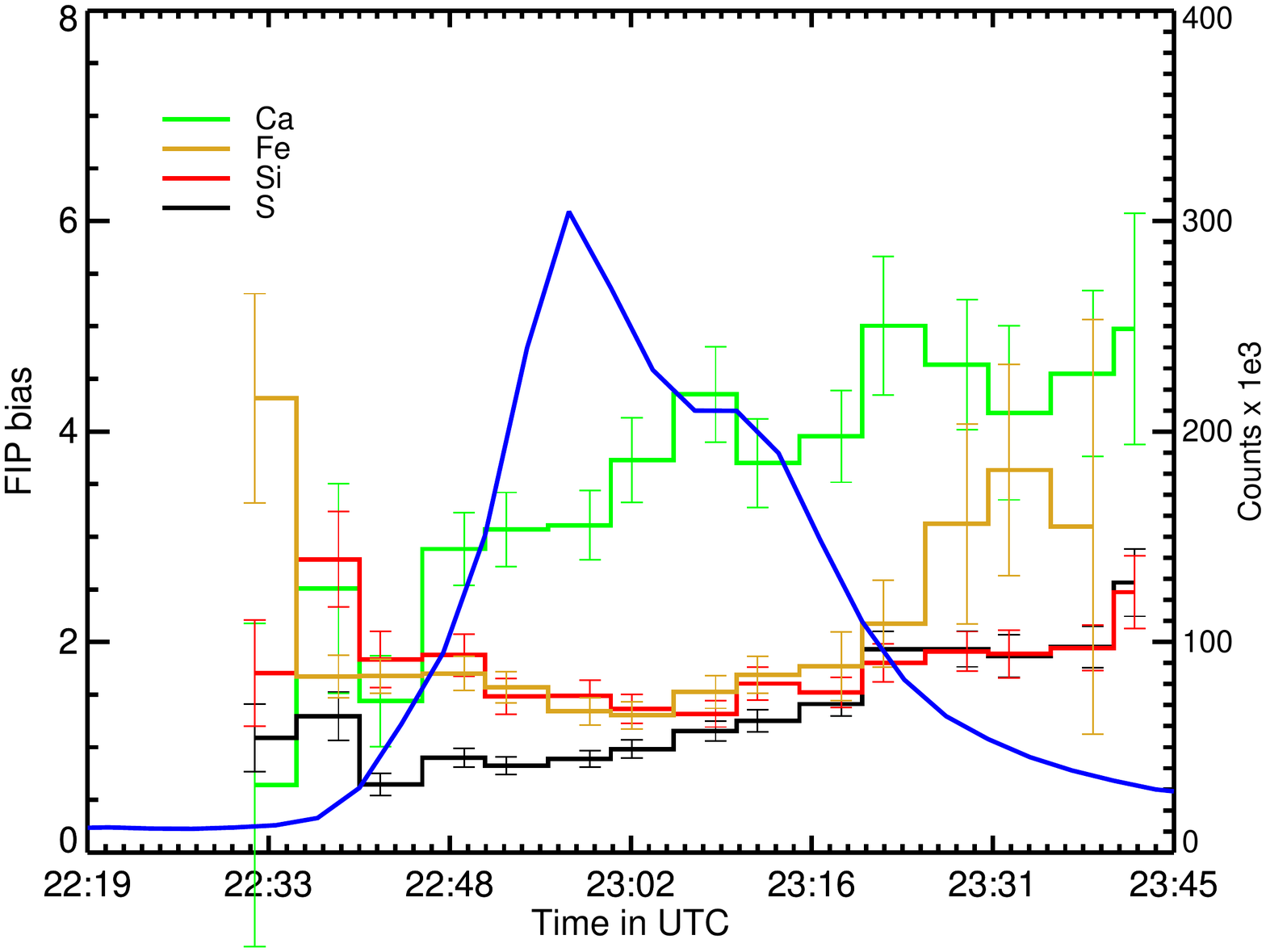}
               \hspace*{-0.03\textwidth}
               \includegraphics[width=0.52\textwidth,clip=]{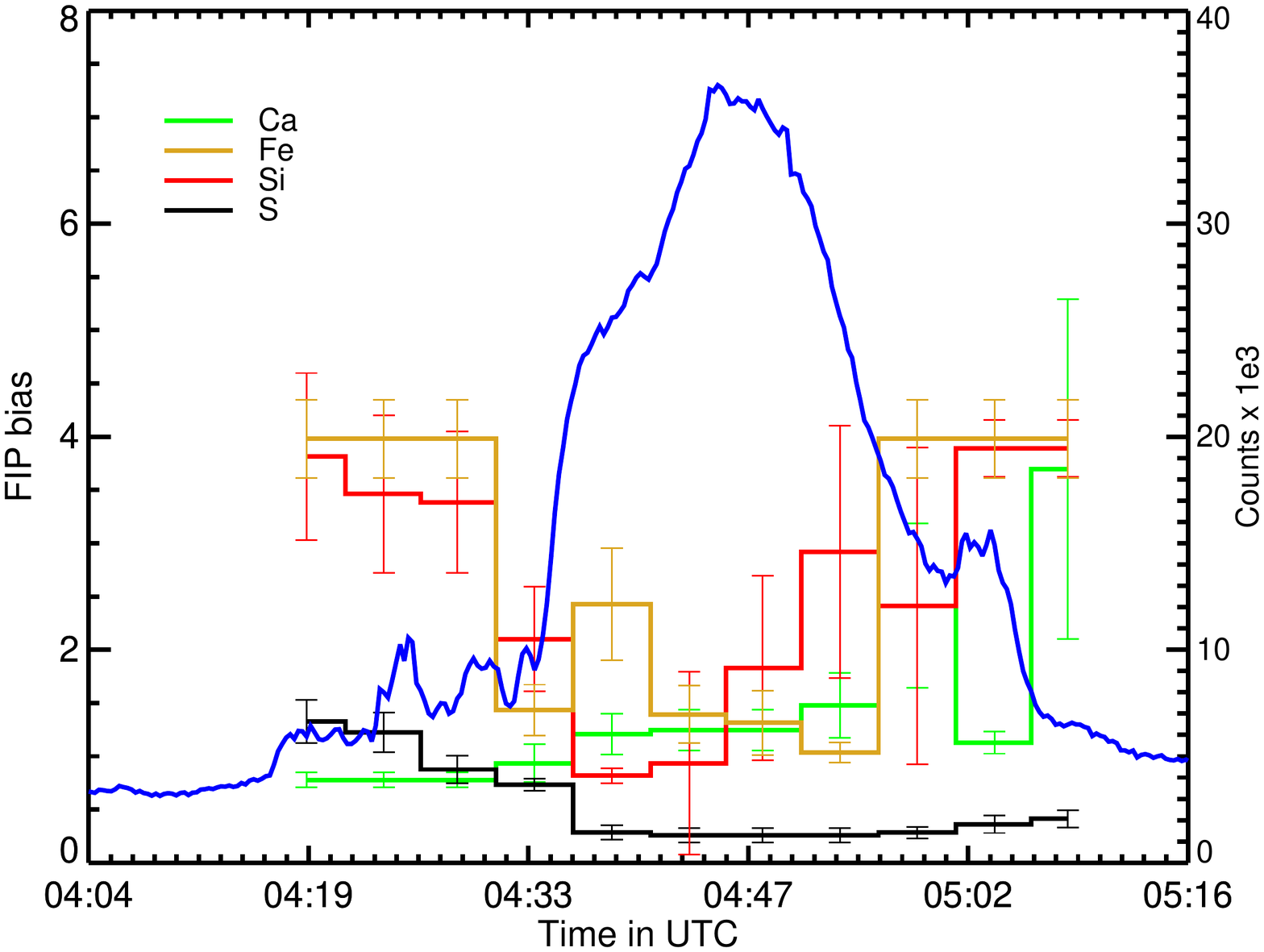}
              }
     \vspace{-0.32\textwidth}   
     \centerline{\Large \bf     
      \hspace{0.15 \textwidth}  \color{red}{(a)}
      \hspace{0.42\textwidth}  \color{red}{(b)}
         \hfill}
     \vspace{0.3\textwidth}    
\caption{FIP bias variation for the flares on (a) 6 January 2014 ( data in Figure 3) (b) 1 May 2004 (data in figure 4) for Ca, Fe, Si and S . The light curve in the 2-10 keV energy range is plotted in blue. The recovery rates are observed to be different. }
\label{all_bias}
\end{figure}

 For flares with sharply defined peaks, the minimum of the FIP bias appears to precede the peak as shown in Figure \ref{all_bias}. Additionally, there is a difference observed in the rate of recovery to coronal values.  

   \par
\begin{figure}[h]    
   \centerline{\hspace*{0.015\textwidth}
               \includegraphics[width=0.4\textwidth,clip=]{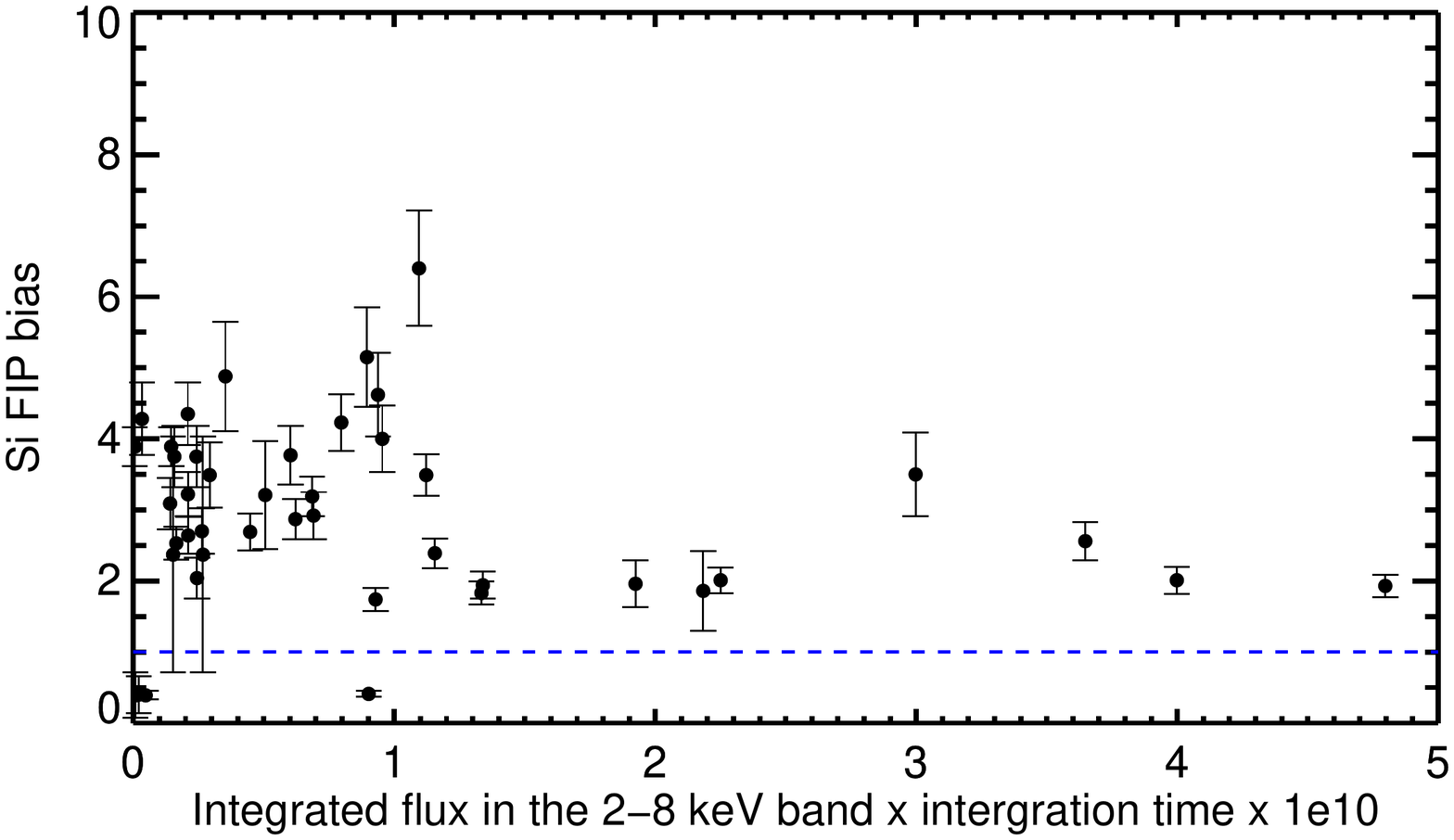}
               \hspace*{-0.03\textwidth}
               \includegraphics[width=0.4\textwidth,clip=]{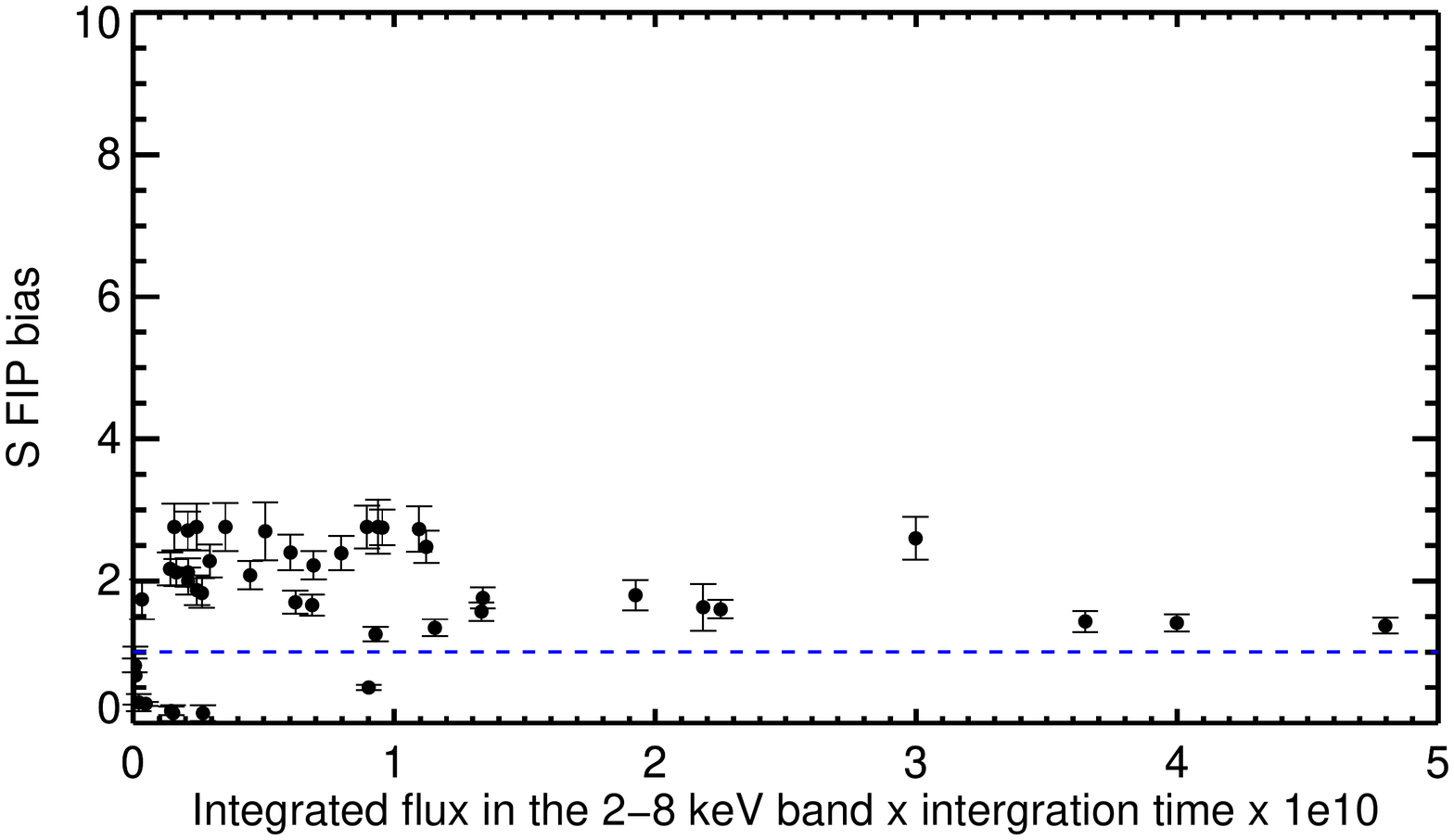}
              }
     \vspace{-0.1\textwidth}   
     \centerline{\Large \bf     
         \hfill}
     \vspace{-0.05\textwidth}    
   \centerline{\hspace*{0.015\textwidth}
               \includegraphics[width=0.4\textwidth,clip=]{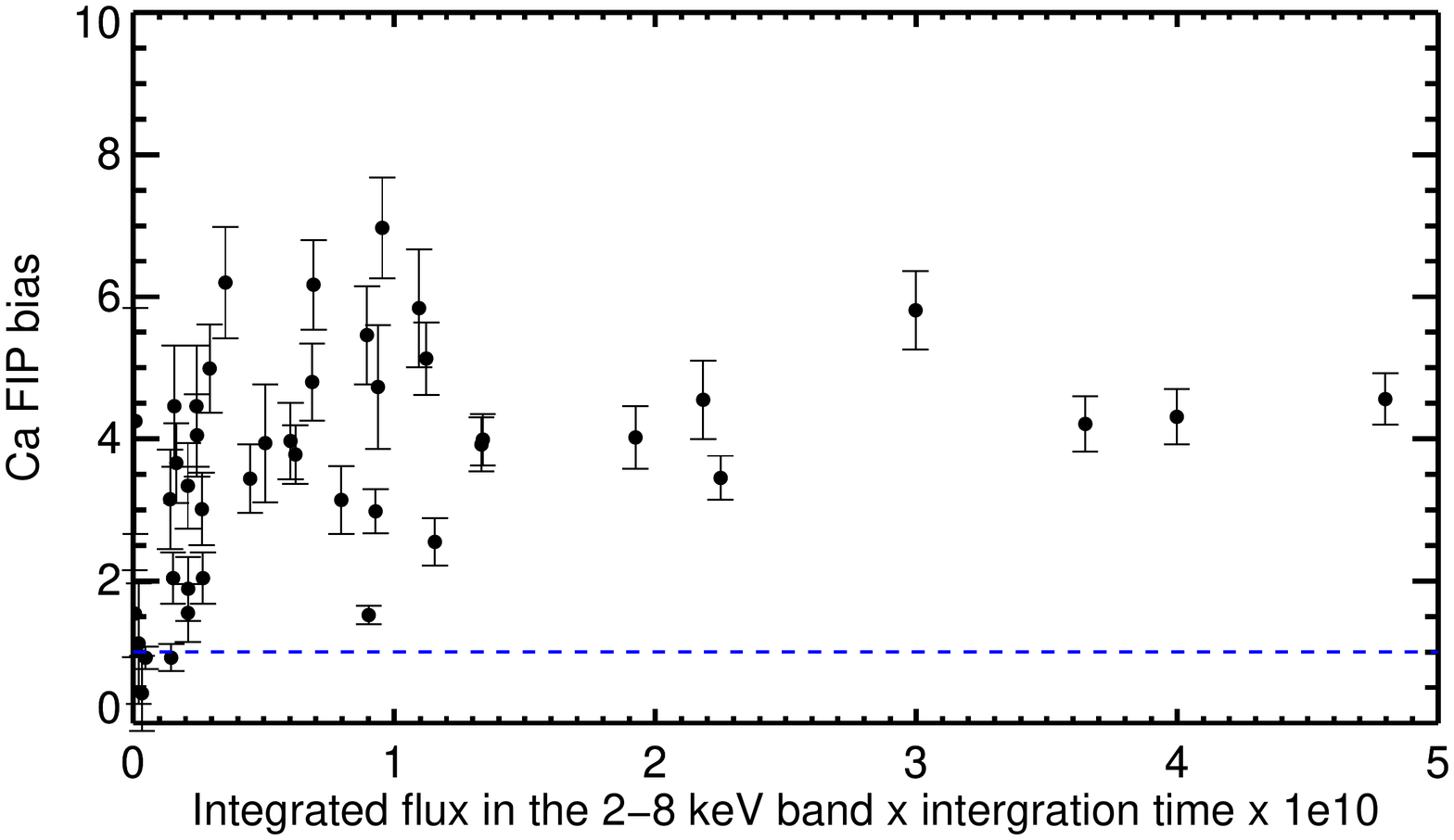}
               \hspace*{-0.03\textwidth}
               \includegraphics[width=0.4\textwidth,clip=]{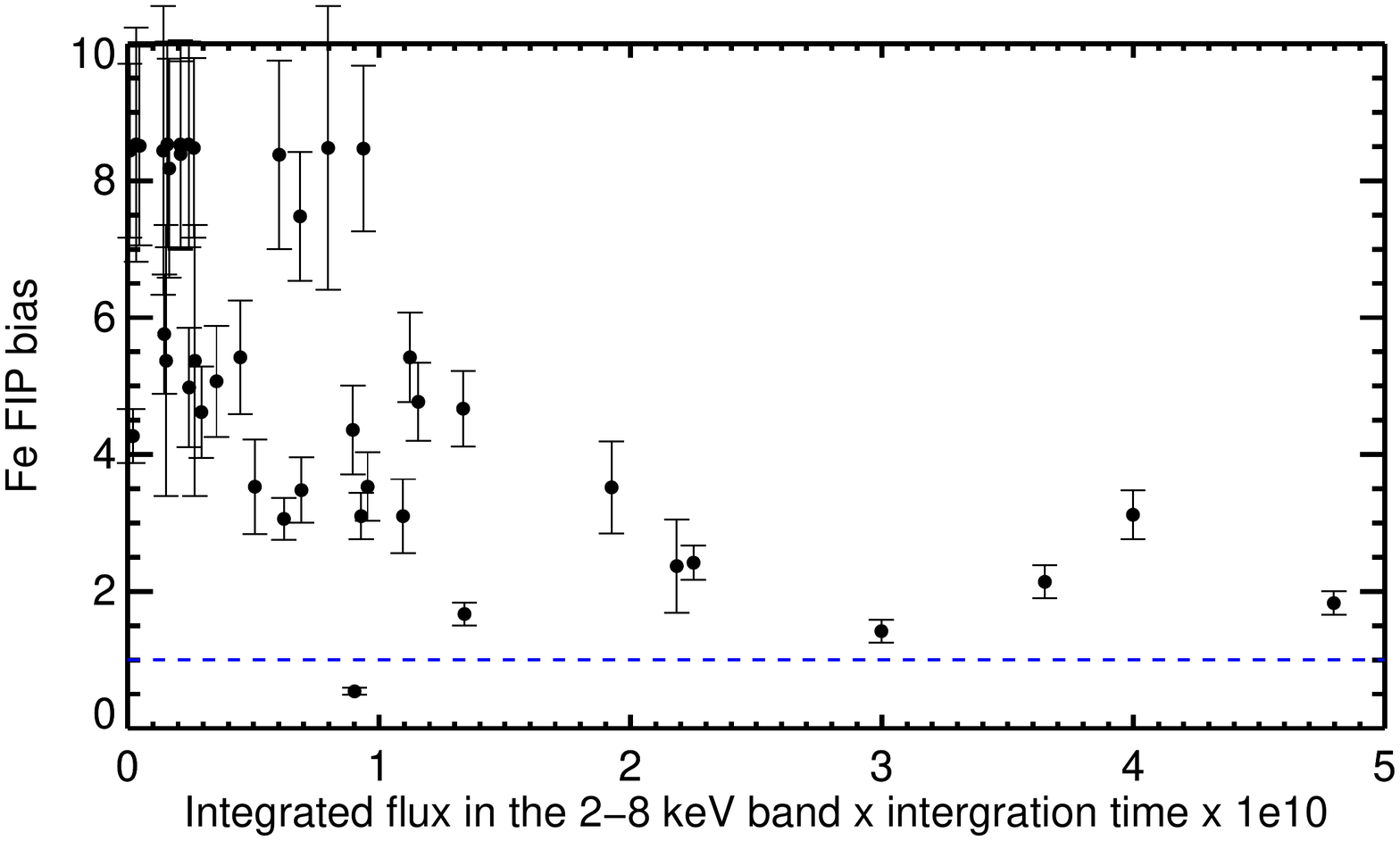}
              }
     \vspace{-0.15\textwidth}   
     \centerline{\Large \bf     
         \hfill}
     \vspace{0.1\textwidth}    
              
\caption{FIP bias variations across flares as a function of the strength of the flare (integrated flux in the 2-10 keV multiplied by the integration time). The spectra during the whole interval of the flare are summed and analysed for 43 flares. The trend suggests a lower FIP bias (and lesser fractionation) for higher intensity flares. A dotted blue line is drawn at FIP bias value=1 which is when coronal abundances are equal to photospheric ones. }
   \label{FIP_flares} 
   \end{figure}

 We also studied FIP bias variations using the full set of 43 flares (Table 2). Data during the whole interval of a flare are summed and analysed, to derive the mean FIP bias for each flare. A dotted  line is drawn at FIP bias value equal to 1 which is when coronal abundances equal  photospheric values.
   The general trend for all elements of lower FIP bias for stronger flares (Figure \ref{FIP_flares}), is consistent with the speculation in \citeauthor{2020ApJ...891..126K} (2020) that more intense flares show larger departures of FIP values from nominal. \citeauthor{2020ApJ...891..126K} (2020) used albedo signal from Earth’s atmosphere during four giant solar flares measured using the X-ray imaging spectrometer on Suzaku. Unlike our work, during the Suzaku observations, the flare spectrum was not continuously monitored to derive the rapidly changing FIP bias values with time, probably due to spacecraft constraints. Instead, bias values were compared from successive orbits with a large time gap of nearly 90 minutes to arrive at their conclusion. Our cumulative data for 43 flares show this trend most clearly for Fe, followed by Si. S is the least fractionated.  The trend is not very evident for Ca even though the detailed analysis of individual flares, shows changes consistent with other low-FIP elements. This could arise from the fact that the data points in Figure \ref{FIP_flares} represent mean FIP bias for the observed flare duration. In addition, Figure \ref{FIP_flares}  shows that Ca has a greater tendency for a quick recovery to pre-flare FIP bias values as the flare crosses its peak. 

 \par There are a few instances of FIP bias falling below photospheric values. \citeauthor{2020ApJ...891..126K} (2020) also reports IFIP values for S and Si during the decay phase. We suggest that in the future this can be investigated in detail using a larger sample of flares and flare classes.

%


\section{Discussion}

\par The coronal FIP bias, first reported by \citeauthor{1963ApJ...137..945P} (1963) and consistently reported in subsequent years by numerous authors is studied during flares using UV and X-ray spectroscopic techniques, through direct sampling  in slow and fast solar wind and during observations of solar energetic events (SEP) in Earth’s vicinity. The degree of enhancement reported in these studies shows a large range. 
\par Explanations for this long standing puzzle of enhancement of low FIP elemental abundance in the solar corona have been attempted by many authors. Currently, the FIP effect is understood as arising from fractionation in the chromosphere and subsequent acceleration of these ions in a force-free field into the corona. Theoretical frameworks are described in \cite{1998SSRv...85..215H}, \cite{2004ApJ...614.1063L}, \cite{2009ApJ...695..954L}, \cite{2012ApJ...744..115L}, \cite{2015LRSP...12....2L}. 
 \citeauthor{2004ApJ...614.1063L} (2004) suggested a ponderomotive force arising from wave refraction in an inhomogenous plasma. In a magnetic plasma, waves with frequencies much lower than the ion cyclotron frequency are refracted to regions of high wave energy density taking along the ions producing the FIP effect. 

\citeauthor{2017ApJ...844..153L} (2017) studied the dependence of fractionation of elements on the origin and propagation of Alfven waves through the chromosphere. The model reproduces the observed fractionations in both open and closed field regions. \citeauthor{Laming_2019} (2019) applied the model to investigate scenarios in source regions of SEPs and the solar wind.     

\par The results from our study here shows the time dependent variation of FIP bias during the evolution of flares. 

Soft X-ray emission in solar flares is dominant from the closed loops that are filled with hot plasma.  One of the ways to understand the systematic reduction in FIP bias during the rising phase of a flare is the following. Assuming that a reconnection process leads to a flare, the pre-flare magnetic field configuration rapidly changes and new channels for upward flow may emerge. Plasma from the upper photosphere fills the coronal loops which are observed in X rays. The fractionation is brought about in the chromosphere by the action of the ponderomotive force. The possible scenarios and parameter dependence of such a model for variations in the degree of fractionation (as observed here) is described in \citeauthor{2017ApJ...844..153L} (2017).  When the upward flow speeds of the plasma is high, it would pass through the chromosphere too quickly to get fractionated. During flares it is likely that the subtle variations in the FIP bias as observed here is indicative of the variations in heating and chromospheric evaporation.

After flare maximum, the field configuration in the lower corona recovers to form new closed shapes in loops that lead to reduced flow rates and higher fractionation.

Though \citeauthor{2017ApJ...844..153L} (2017) model is more suitable for quiet Sun conditions, the FIP bias decrease during a flare observed in this work can be attributed to the presence of an open field configuration during a reconnection process in active regions.

\section{Summary}
The evolution of absolute elemental abundances of Si, S, Ca, and Fe in the flare plasma has been determined from soft X-ray spectroscopy. We show that during flare peaks, the abundance of low FIP elements significantly reduces to photospheric values. Short time-scale variations in the abundance values observed in flares provide a framework  to model the wave propagation and plasma  heating during disturbed solar conditions. Future measurements extending down to 0.5 keV with high resolution spectrometers along with chromospheric velocity variation and magnetic field measurements during the evolution of a flare would provide valuable additional information to further constrain the model.

\begin{acks}
 We thank Brian Dennis, Richard Schwartz, Kim Tolbert at GSFC and Vinay Kashyap at CFA for useful discussions under the Indo-US Science and Technology Forum. We extend our gratitude to IUSSTF grant JC-24-2015 for funding this research. We also very much appreciate the comments and suggestions from the anonymous reviewer that has helped us improve the paper.
\end{acks}

Disclosure of Potential Conflicts of Interest: The authors declare that there are no conflicts of interest.


\bibliographystyle{spr-mp-sola}
\bibliography{XSM_paper_v6}  

\IfFileExists{\jobname.bbl}{} {\typeout{}
\typeout{****************************************************}
\typeout{****************************************************}
\typeout{** Please run "bibtex \jobname" to obtain} \typeout{**
the bibliography and then re-run LaTeX} \typeout{** twice to fix
the references !}
\typeout{****************************************************}
\typeout{****************************************************}
\typeout{}}

\end{article} 

\end{document}